\documentclass[journal,12pt,draftclsnofoot,onecolumn,english]{IEEEtran}
\usepackage[T1]{fontenc}
\usepackage{url}
\usepackage{ifthen}
\usepackage{cite}

\usepackage{amsthm}
\usepackage{epsfig}
\usepackage{times}
\usepackage{float}
\usepackage{afterpage}
\usepackage{sansmath}
\usepackage{amsmath}
\usepackage{amstext,cite}
\usepackage{amssymb,bm}
\usepackage{latexsym}
\usepackage{color}
\usepackage{graphicx}
\usepackage[center]{caption}
\usepackage{pstricks}
\usepackage{caption}
\usepackage{subcaption}
\usepackage{booktabs}
\usepackage{multicol}
\usepackage{lipsum}
\usepackage{hyperref}
\usepackage{aecompl}
\usepackage{mathrsfs}
\usepackage{blkarray}
\usepackage{arydshln}
\usepackage{algorithm}
\usepackage{algpseudocode}
\usepackage{cite}
\usepackage{soul}
\usepackage{cancel}
\usepackage{amssymb}
\usepackage{amsthm} 
\newtheorem{proposition}{Proposition}  % 定义 proposition 环境
\newcommand{\vdashline}{%
  \tikz[baseline]{\draw[dashed] (0,-0.8ex) -- (0,1.8ex);}%
}
\usepackage{changes}
\definechangesauthor[color=red,name={Mingyue Ji}]{MJ}

\allowdisplaybreaks

\newcommand{\insertxfig}[4]{
\begin{figure}[htbp]
\begin{center}
\leavevmode \centerline{\resizebox{#4\textwidth}{!}{\input
#1.pstex_t}}
\caption{#2} \label{#3}
\end{center}
\end{fontsize}}

\long\def\comment#1{}

\newcommand{\sv}{{\mathbf s}}

\newcommand{\Ac}{{\mathcal A}}
\newcommand{\Bc}{{\mathcal B}}
\newcommand{\Cc}{{\mathcal C}}
\newcommand{\Dc}{{\mathcal D}}

\newcommand{\Fc}{{\mathcal F}}
\newcommand{\Gc}{{\mathcal G}}

\newcommand{\Ic}{{\mathcal I}}

\newcommand{\Qc}{{\mathcal Q}}

\newcommand{\Sc}{{\mathcal S}}

\newcommand{\Uc}{{\mathcal U}}

\newcommand{\Vc}{{\mathcal V}}

\newcommand{\Zc}{{\mathcal Z}}

\newcommand{\asf}{{\mathsf a}}

\newcommand{\csf}{{\mathsf c}}

\newcommand{\rsf}{{\mathsf r}}

\newcommand{\Csf}{{\mathsf C}}

\newcommand{\Ksf}{{\mathsf K}}
\newcommand{\Lsf}{{\mathsf L}}
\newcommand{\Msf}{{\mathsf M}}
\newcommand{\Nsf}{{\mathsf N}}

\newcommand{\Tsf}{{\mathsf T}}

\renewcommand{\det}{{\hbox{det}}}

\newtheorem{thm}{Theorem}%[section]

\newtheorem{lem}{Lemma}

\newtheorem{example}{Example}

\newtheorem{constraint}{Constraint}

\providecommand{\definitionname}{Definition}

\usepackage{tikz}
\usetikzlibrary{calc}
\usetikzlibrary{fit}
\usepackage[nodisplayskipstretch]{setspace} %\setstretch{1.5}

\begin{document}
\title{Distributed Linearly Separable Computation with Arbitrary Heterogeneous Data Assignment} 
% Many authors with many affiliations:
\author{
Ziting~Zhang, 
Kai~Wan,~\IEEEmembership{Member,~IEEE,} 
Minquan~Cheng,~\IEEEmembership{Member,~IEEE,}  
Shuo~Shao,~\IEEEmembership{Member,~IEEE,}
and~Giuseppe Caire,~\IEEEmembership{Fellow,~IEEE}
\thanks{
Z.~Zhang and K.~Wan are with the School of Electronic Information and Communications,
Huazhong University of Science and Technology, 430074  Wuhan, China  (e-mail: \{ziting\_zhang,kai\_wan\}@hust.edu.cn). %The work of Z.~Zhang and K.~Wan was  funded by the National Natural Science Foundation of China under Grant  12141107, the Key Research and Development Program of Wuhan under Grant 2024050702030100, and Wuhan ``Chen Guang'' Program under Grant 2024040801020211. 
}
\thanks{M.~Cheng is with the Key Laboratory of Education Blockchain and
Intelligent Technology, Ministry of Education, and Guangxi Key Laboratory of
Multi-Source Information Mining and Security, Guangxi Normal University,
541004 Guilin, China (e-mail: chengqinshi@hotmail.com).}
\thanks{S.~Shao is with the Department of System Science, University of
Shanghai for Science and Technology, 200093 Shanghai, China (e-mail: shuoshao@usst.edu.cn).}
\thanks{G.~Caire is with the Electrical Engineering and Computer Science Department, Technische Universit\"at Berlin, 10587 Berlin, Germany (e-mail: caire@tu-berlin.de). %The work of G. Caire was supported by the Gottfried Wilhelm Leibniz-Preis 2021 of the German Science Foundation (DFG).
}
}

\maketitle
\begin{abstract}
Distributed linearly separable computation is a fundamental problem in large-scale distributed systems, requiring the computation of  linearly separable functions over different datasets  across distributed workers. This paper studies a heterogeneous distributed linearly separable computation problem, including one master and  $\Nsf$ distributed workers. The linearly separable task function involves $\Ksf_\csf$ linear combinations of $\Ksf$ messages, where each message is a function of one dataset. Distinguished from the existing homogeneous settings that assume each worker holds the same number of datasets, where the data assignment is carefully designed and controlled by the data center (e.g., the cyclic assignment), we consider a more general  setting with arbitrary heterogeneous data assignment across workers, where `arbitrary' means that the data assignment is given in advance and `heterogeneous' means that the workers may hold different numbers of datasets. 
Our objective is to characterize the fundamental tradeoff between the computable dimension of the task function 
and the communication cost under arbitrary heterogeneous data assignment. %We show that the problems of minimizing the communication cost for a given computable dimension and maximizing the computable dimension under a fixed communication cost are equivalent. In our setting, we focus on maximizing the computable dimension subject to a fixed communication cost. 
Under the constraint of integer communication costs, for arbitrary heterogeneous data assignment, we propose a universal computing scheme   and  a universal converse bound by characterizing the structure of data assignment, where they coincide under some parameter regimes. We then extend the proposed computing scheme and converse bound to the case of fractional communication costs.

%without subpacketization, which corresponds to integer communication costs. The scheme reduces communication cost by decomposing the computation into $\Ksf_\csf$ distributed processes, each associated with a single demand dimension, and can be extended to support arbitrary fractional communication costs. We further derive a converse bound on the maximum computable dimension, which coincides with the achievable scheme under some parameter regimes.
\end{abstract}


\section{Introduction}
Distributed computing systems, such as MapReduce \cite{dean2008mapreduce} and Spark \cite{zaharia2010spark}, have emerged as essential solutions for large-scale computations on massive datasets \cite{dean2012large}. A common strategy is to divide computational tasks into smaller subtasks and distribute them across multiple computing nodes for parallel execution. However, these systems face significant challenges, including high communication costs \cite{li2014communication}  and the impact of stragglers \cite{dean2013tail}.
% However, large-scale distributed computing faces several challenges, in particular (i) the high communication costs associated with transferring large amounts of data over limited communication bandwidth \cite{li2014communication}; and (ii) the impact of stragglers (such as unstable network connections or delayed transmissions), which severely degrade overall computation time \cite{dean2013tail}. 
Coding techniques have been proven to address these challenges \cite{lee2017speeding,li2015coded,yu2019lagrange,yu2017polynomial,halbawi2018improving}.

Since the introduction of coded distributed computation based on coded storage~\cite{lee2017speeding}, significant progress has been made in designing approaches that balance the computation load and communication cost under robustness constraints. 
Various information-theoretic coded distributed computation approaches have been proposed, which could generally be classified into two
frameworks according to   data assignment:\begin{itemize}
\item Coded data assignment. Most of these approaches focus on three classes of computational tasks: distributed matrix-vector multiplication\cite{ramamoorthy2019universally,das2019distributed,qiu2024coded}, distributed matrix-matrix multiplication\cite{lee2017high,dutta2019optimal,yu2017polynomial,son2023coded}, and multivariate polynomial evaluation\cite{yu2019lagrange,zhu2024generalized}. As
in coded distributed matrix multiplication, the input matrices are first partitioned into submatrices, after which classical error-correcting codes, such as minimal distance separable (MDS)  or polynomial codes, are employed to form linear combinations of these submatrices. These encoded submatrices are then assigned to different computing nodes for computation. A key limitation is that coded data assignment typically relies on centralized preprocessing at a data center. This centralized approach, however, incurs prohibitive overhead for complex iterative computations such as gradient descent, limiting scalability and practical applicability in large-scale systems.

\item Uncoded data assignment. In this framework, computations are performed directly on uncoded datasets allocated to the nodes. As in distributed gradient coding \cite{gradiencoding,ye2018communication} and distributed linear transform \cite{dutta2016short}, when the master aims to compute a function of $\Ksf$ datasets $(D_1,\ldots,D_\Ksf)$ across $\Nsf$ distributed workers, each worker can only perform computations on the subset of original datasets assigned to it.  As a generalization of distributed gradient coding and distributed linear transform, 
distributed linearly separable computation was originally proposed in \cite{wan2021distributed}. The task function contains $\Ksf_\csf$ linear combinations of messages; each message is a separable function (such as gradient operation) of one dataset. %and contains $\Lsf$ symbols on a sufficiently large alphabet. 
 This framework operates in three phases: assignment, computing, and decoding. During the
assignment phase, the master assigns parts of $\Ksf$ datasets to workers in an uncoded manner, and the number of assigned datasets to each node is defined as the computation cost by this node. In the homogeneous assignment setting, the computation cost of each node is the same, equal to $\Msf$. During the following computing phase, each worker computes an intermediate
message for each assigned dataset and then transmits a coded version of these intermediate messages to the master. Finally, during the decoding phase, the master can recover the task function from the responses of any
$\Nsf_\rsf$ workers with high probability, ensuring the system can tolerate up to $\Nsf-\Nsf_\rsf$ stragglers. Given the computation cost constraint, the objective is to minimize the number of transmissions that should be received by the master in the worst case. 
\end{itemize}

\subsection{Overview of distributed linearly separable computation}
The information theoretic results of the distributed linearly separable computation with homogeneous computation cost  are summarized as follows:
\begin{itemize}
	\item $\Ksf_\csf=1$. In this case, the problem reduces to the
	distributed gradient coding problem. When the computation cost is minimum, i.e., $\Msf=\frac{\Ksf}{\Nsf}(\Nsf-\Nsf_\rsf+1)$, the gradient coding scheme proposed in \cite{gradiencoding} achieves the optimal communication cost of $\Nsf_\rsf$. Subsequently, \cite{ye2018communication,cao2021adaptive} characterized the optimal tradeoff between the computation and communication costs under the linear coding constraint.
	%\item The optimal communication cost under the cyclic assignment was characterized in \cite{wan2021distributed} for the case of minimum computation cost $\Msf=\frac{\Ksf}{\Nsf}(\Nsf-\Nsf_\rsf+1)$.
	\item $\Ksf_\csf\geq 1$ and $\Msf=\frac{\Ksf}{\Nsf}(\Nsf-\Nsf_\rsf+1)$. A computing scheme was proposed in \cite{wan2021distributed},  which is exactly optimal
when $\Nsf = \Ksf$, and is optimal under the constraint of the cyclic assignment. 
	\item $\Ksf_\csf\geq 1$ and $\Msf\geq \frac{\Ksf}{\Nsf}(\Nsf-\Nsf_\rsf+1)$.
	 \cite{wan2021tradeoff,huang2023fundamental} proposed computing schemes achieving the order-optimal communication cost within a factor of $2$ under the cyclic assignment constraint.
	%\item The optimal communication cost under the cyclic assignment was characterized in \cite{wan2021distributed} for the case of minimum computation cost $\Msf=\frac{\Ksf}{\Nsf}(\Nsf-\Nsf_\rsf+1)$.
\end{itemize}

Another research direction in distributed linearly separable computation is based on coding theory. Different from the information theoretic formulation, which allows one to divide each message into sub-messages and express the task function as linear combinations of sub-messages, the  coding-theory-based model does not consider message division; thus, the communication cost for each worker is an integer.
\iffalse 
Then the problem is formulated as 
%Existing studies on distributed linearly separable computation assume integer communication costs and aim to characterize the fundamental trade-off between computation cost and communication cost. A common approach models the problem as 
 a sparse matrix factorization $\mathbf{F} = \mathbf{D}\mathbf{E}$, where the objective is to minimize the number of nonzero entries in the encoding or decoding matrices. 
 \fi 
 In particular,
 by reformulating the problem as a sparse matrix factorization problem on $\mathbf{F} = \mathbf{D}\mathbf{E}$ where $\mathbf{F}$ represents the task function, $\mathbf{D}$ represents the decoding matrix of the master, and $\mathbf{E}$ represents the concatenated encoding matrices of the workers, several schemes have been proposed:
\begin{itemize}
    \item Multi-server Matrix Factorization.
    The works in \cite{khalesi2023multi}, \cite{khalesi2024perfect} provided the optimal computation cost bounds for the error-free case, and \cite{khalesi2025tessellated} further explored lower bounds on both communication and computation costs in the lossy scenario.
    \item Single-server Matrix Factorization. In \cite{namboodiri2025fundamental}, the authors characterized the fundamental tradeoff between communication cost and computation cost, which has been proven to be either exactly optimal or order optimal within a factor of $3$.  
\end{itemize}

In many real-world distributed learning systems, the data assignment to workers cannot be carefully designed in advance. A representative example is the Mixture-of-Experts (MoE) architecture, which has become a core component of modern large-scale language models such as DeepSeek\cite{rajbhandari2022deepspeed,lin2024moe}. In the MoE system, input data are dynamically routed to a subset of experts through a gating mechanism, and each expert is typically specialized for certain data types or tasks. Due to such specialization, as well as practical constraints such as heterogeneous hardware capacities and memory limits, each expert can only process the data assigned to them, resulting in inherently fixed and asymmetric data assignment across experts. To address arbitrary heterogeneous data assignment in this setting, a distributed gradient coding framework (i.e., the case $\Ksf_\csf=1$ in distributed linearly separable computation) was introduced in \cite{jahani2021optimal}. By defining the communication cost as the maximum transmission load among all nodes, a universal linear coding scheme was proposed to achieve the minimum communication cost. Specifically, it is proven that the optimal communication cost depends solely on the minimum replication factor $r$ of the data partitions. For a system with $s$ stragglers and $a$ adversarial nodes, it is shown that the minimum normalized communication cost is $\frac{1}{r - s - 2a}$, implying that the overall system performance is bottlenecked by the least-replicated data partition.

Beyond single-task gradient coding (i.e., $\Ksf_{\rm c}=1$), the authors in \cite{cheng2025novel} study distributed linearly separable computation for $\Ksf_{\rm c}\geq 1$. They propose a coding scheme that achieves the optimal tradeoff between computation cost and communication cost by selecting a virtual demand matrix satisfying the MDS property, where the encoding vectors for each worker can be directly obtained from the left null space of the corresponding submatrix. However, this scheme holds only under a specific condition: for any positive integer $t \in [\Nsf]$, $\left| \bigcap_{n \in [t]} \overline{\Sc_n}^{\Nsf-1} \right| \leq (\Nsf-1)(\sum_{n\in[\Nsf]}\Csf_n - \sum_{n \in [t]} \Csf_n)$, where $\overline{\Sc_n}^{\Nsf-1} = \bigcup_{i\in[0:\Nsf-2]}(\overline{\Sc_n}+i\Ksf)$ and $\Csf_n=\frac{\Ksf+(\Nsf-1)|\Sc_n|-\sum_{n^\prime\in[\Nsf]}|\Sc_{n^\prime}|}{\Nsf-1}$. Here, $\overline{\Sc_n}$ 
denotes the set of datasets not assigned to worker $n$, and 
$\Csf_n$ represents to the transmission of worker $n$. Intuitively, this condition requires that the heterogeneity in data assignment does not overly restrict the available encoding degrees of freedom across workers, which are necessary to ensure the linear independence of the encoding vectors.

The existing studies have made substantial progress in distributed linearly separable computation, particularly in optimizing communication-computation tradeoffs. However, an open problem still remains in the case of $\Ksf_{\rm c}>1$ with arbitrary heterogeneous data assignment.

  \subsection{Main contributions}
  Our framework is built on the coding-theory-based formulation, established in prior work \cite{khalesi2023multi, khalesi2024perfect, khalesi2025tessellated, khalesi2025lossless,namboodiri2025fundamental}. %We operate under the no-subpacketization constraint, which yields integer communication costs, and further extend the framework to the case of fractional communication costs. The key novelty of our framework is that, 
  Unlike conventional schemes based on symmetric cyclic or disjoint data assignment, we consider  the distributed linearly separable computation problem with $\Ksf_{\rm c}\geq 1$ and  arbitrary heterogeneous data assignment.
  %  arbitrary heterogeneous data assignment and naturally extends to multi-task settings with multiple linearly separable demands, providing a universal coding scheme for arbitrary heterogeneous data assignment.
  Note that the considered problem is also a generalization of distributed gradient coding problem with 
arbitrary heterogeneous data assignment  in 
%In this paper, we generalize the problem in 
\cite{jahani2021optimal}, by considering distributed linearly separable computation with 
arbitrary heterogeneous data assignment for the case of $\Ksf_\csf\geq 1$. Our objective is to characterize the fundamental tradeoff between the communication cost and the 
maximum computable dimension $\Ksf_\csf$. %under a fixed communication cost. 
Compared to the work in~\cite{cheng2025novel}, we aim to propose universal achievable and converse bounds 
for arbitrary heterogeneous data assignment,   without any constraint on the data assignment nor  on the system parameters.
More precisely, our main contributions are as follows.  
\begin{itemize}
    \item {\bf Integer communication costs.}
    As in~\cite{khalesi2023multi, khalesi2024perfect, khalesi2025tessellated, khalesi2025lossless,namboodiri2025fundamental}, we first consider the integer communication costs by disallowing the division/subpacketization on the messages during the linear coding at the transmission side.  
   For arbitrary heterogeneous data assignment, we derive a universal converse bound on the maximum computable dimension $\Ksf_\csf$ under any fixed communication cost, and also propose a universal coded computing scheme, by leveraging the combinatorial structure of the data assignment. The converse and achievable bounds coincide under some parameter regimes.%\footnote{\label{foot:novelty} In the proposed coded computing scheme, we construct a matrix via two nested null-space operations and prove its full-rank property based on a novel application of the Schwartz--Zippel Lemma. This construction provides a key theoretical guarantee for our achievable scheme.} 
   %without subpacketization, accommodating arbitrary heterogeneous data assignment; in this setting, the scheme corresponds to integer communication costs, and it can be further extended to the case of fractional communication costs.

    \item {\bf Fractional communication costs.} 
  We then extend the model to fractional communication costs, allowing each dataset to be partitioned into $q$ non-overlapping and equal-length sub-messages. Each worker transmits at most $p$ linear combinations of these sub-messages so that the master can recover $q \Ksf_\csf$ linear combinations of the sub-messages with high probability. By leveraging the main strategies of the case of integer communication costs,
     we also obtain both converse and achievable bounds on the tradeoff between  the communication cost $p/q$ and the computable dimension $\Ksf_{\rm c}$.
    % \item  {\red Based on experimental results conducted on}
\end{itemize}

%    \item {\bf Arbitrary communication costs.} When $\Csf=\frac{p}{q}\Lsf$ for any $p,q\in\mathbb{N}^+$, we extend both the converse bound and the proposed scheme to  formulate a compression strategy.

\subsection{Notation convention}
%We use the following notation convention.
Sets are denoted using calligraphic symbols.
Vectors and matrices are represented in bold.
System parameters are indicated in sans-serif font.
The notation $[n]$ denotes the set $\{ 1, 2, \ldots, n \}$. The matrix $[a;b]$ is expressed in a format similar to Matlab, equivalent to $\begin{bmatrix}
     a\\ 
     b
 \end{bmatrix} $. $\mathbf{M}^{-1}$ represents the inverse of matrix $\mathbf{M}$. $\text{rank}(\mathbf{M})$ represents the rank of matrix $\mathbf{M}$.
For a set $\Sc$, we denote the $i^{\text{th}}$ smallest element by $\Sc(i)$; $\mathbf{M}(\Sc,.)$ and $\mathbf{M}(.,\Sc)$ 
represent the sub-matrices of $\mathbf{M}$  composed of the
rows  with indices in $\Sc$ and the   columns  with indices in $\Sc$, respectively.  $\mathbf{M}(\{i\},\{j\})$ denotes the element in the $i^{\text{th}}$ row and $j^{\text{th}}$ column of the matrix $\mathbf{M}$. For a set $\Ac$ and an integer $i$, we denote $\Ac+i:=\{a+i|a\in\Ac\}$, i.e., the set obtained by adding $i$
 to each element in $\Ac$.
\begin{figure}  % H 让图片固定在当前位置
    \centering
    \includegraphics[width=0.7\textwidth]{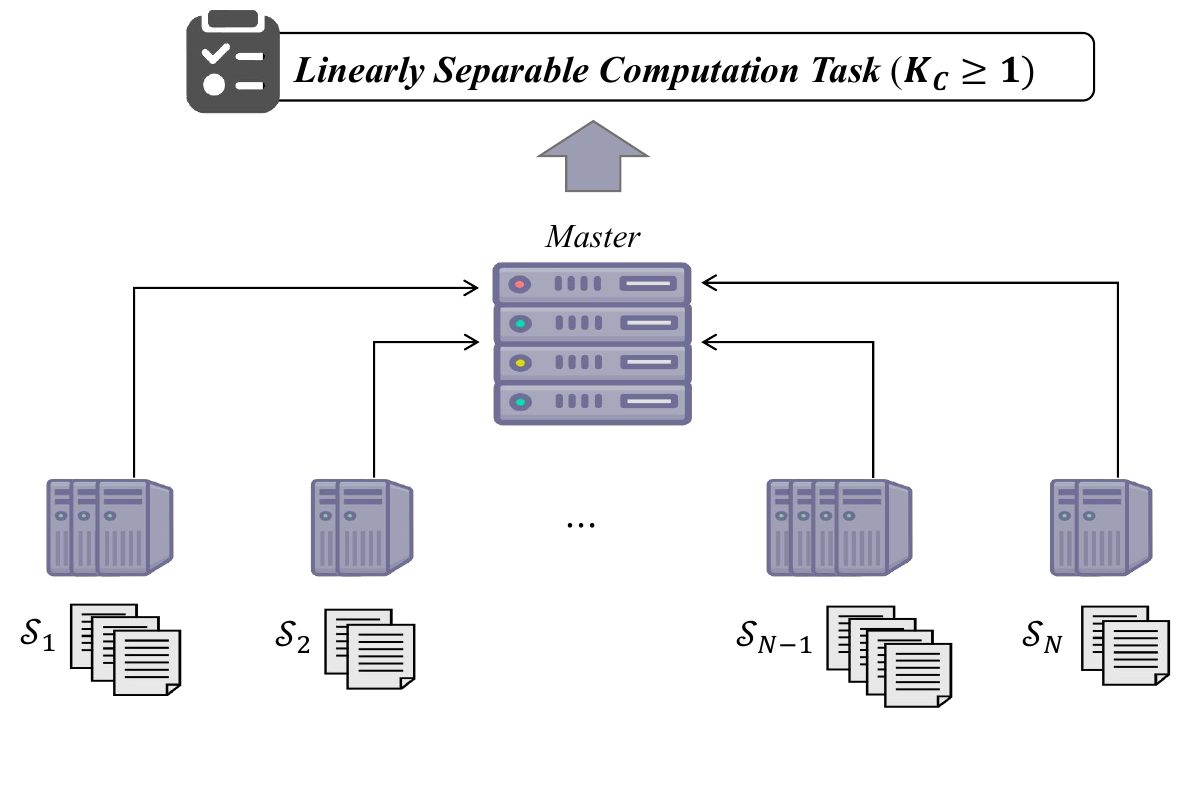}  
    \caption{A $(\Ksf,\Nsf,\Csf)$ distributed linearly computation system with arbitrary heterogeneous data assignment.}  % 图片标题
    \label{fig:example}  % 交叉引用标签
\end{figure}
\section{System Model and Related Results}
\subsection{Problem formulation}
\label{sub:formulation}
We consider a $(\Ksf,\Nsf,\Csf)$ distributed linearly separable computation problem with arbitrary heterogeneous data assignment, 
 consisting of one master and $\Nsf$ workers, as illustrated in Fig.~\ref{fig:example}. As in~\cite{wan2021distributed}, the master aims to compute a linearly separable function $f$ over $\Ksf$ datasets, denoted by $\Dc=\{D_1,\ldots, D_\Ksf\}$.
    The function $f$ is expressed as:
    $$f(D_1,\ldots, D_\Ksf)=g(f_1(D_1),\ldots,f_\Ksf(D_\Ksf))=g(W_1,\ldots,W_\Ksf),$$
    where each $f_k(D_k)$ produces an outcome $W_k$ (a string of 
    $\Lsf$ symbols over a sufficiently large alphabet $\mathbb{F}_{\asf}$) on dataset $D_k$ for  $k\in[\Ksf]$.\footnote{\label{foot:real}The proposed schemes can be applied to the field of real numbers. We assume the finite field for the sake of deriving converse bounds.}
     These intermediate outcomes (referred to as messages) $W_1,\ldots,W_\Ksf$ are then linearly combined by the function 
    $g(.)$ to obtain $\Ksf_\csf$ linear combinations with uniformly i.i.d. coefficients over $\mathbb{F}_{\asf}$;  $g(W_1,\ldots,W_\Ksf)$ can be written as \begin{align}g(W_1,\ldots,W_\Ksf)=\begin{bmatrix}
        F_1\\
        \vdots\\
        F_{\Ksf_\csf}
\end{bmatrix}=\mathbf{F}_1\begin{bmatrix}
        W_1\\
         \vdots\\
        W_\Ksf
    \end{bmatrix},\label{the task}\end{align}
     where $\mathbf{F}_1$ is the coefficient matrix of dimension $\Ksf_\csf\times\Ksf$. 
     Let $\mathbf{W}=[W_1;\cdots;W_\Ksf]$, the master should recover $\mathbf{F}_1\mathbf{W}$ from the messages of  $\Nsf$ workers. %This redundancy ensures that even if up to $s$ workers fail, remaining nodes can still transmit the necessary data for completing the computation.  
     
     Consider a data assignment matrix $\mathbf{A}$ of dimension $\Nsf\times\Ksf$, representing the assignment of $\Ksf$ datasets across $\Nsf$ workers. If dataset $D_k$ is assigned to worker $n$, we let $\mathbf{A}(n,k)=*$; otherwise, we let $\mathbf{A}(n,k)=0$. 
\iffalse      
     The entries in $\mathbf{A}$ are defined as follows:
    \begin{itemize}
        \item If the dataset $D_k$ is assigned to worker $n$, we let $\mathbf{A}(n,k)=*$.
         \item If the dataset $D_k$ is not assigned to worker $n$, we let $\mathbf{A}(n,k)=0$.
    \end{itemize}
\fi      
   Each dataset $D_k$, where $k\in[\Ksf]$, is assigned to a subset of workers $\Cc_k$. To achieve lossless calculation, each dataset $D_k$ where $k\in[\Ksf]$ must be assigned to at least one worker. The data assignment matrix $\mathbf{A}$ should satisfy that 
   \begin{align}
  1\leq|\Cc_k|\leq\Nsf, \ \forall k\in [\Ksf]. \label{eq:data assignment constraint}
   \end{align}
   Let $\overline{\Cc_{k}} \subseteq [\Nsf]$ denote the set of workers that are not assigned dataset $D_k$. In addition, 
  the set of datasets assigned to each worker $n\in[\Nsf]$ is denoted by $\Sc_n$, where $0<|\Sc_n|\leq\Ksf$ and $\Sc_n\subseteq[\Ksf]$.  Let $\overline{\Sc_{n}}$ denote the set of datasets that are not assigned to worker $n$.

Next, we consider the following two phases in our problem:
    
\text{\textit{1) Computing Phase:}} Each worker $n\in[\Nsf]$ first computes intermediate results $\{W_k:k\in\Sc_n\}$ from its assigned datasets, and then applies linear encoding to these results. Specifically, each worker $n\in[\Nsf]$ transmits $\Csf_n$ linearly encoded messages, each constructed as\footnote{We adopt the standard no-subpacketization assumption, under which datasets are not partitioned into smaller subsets.}
\begin{align}
X_{n,i} = \sum_{k\in\Sc_n}e_{n,i,k}W_k\label{X_n,i}, \forall i\in[\Csf_n],
\end{align}
where $e_{n,i,k}$ denotes the encoding coefficient associated with $W_k$ in the $i$-th encoded message. Each worker $n$ sends $X_n=[X_{n,1},X_{n,2},\ldots,X_{n,\Csf_n}]^{\Tsf}$ to the master, which is
\begin{align}
    \begin{bmatrix}
        X_{n,1}\\
        X_{n,2}\\
        \vdots\\
        X_{n,\Csf_n}
    \end{bmatrix}=\begin{bmatrix}
        e_{n,1,1}&e_{n,1,2}&\ldots&e_{n,1,\Ksf}\\
         e_{n,2,1}&e_{n,2,2}&\ldots&e_{n,2,\Ksf}\\
         \vdots& \vdots& \ddots& \vdots\\
          e_{n,\Csf_n,1}&e_{n,\Csf_n,2}&\ldots&e_{n,\Csf_n,\Ksf}
    \end{bmatrix}\begin{bmatrix}
        W_1\\
         \vdots\\
        W_\Ksf
    \end{bmatrix}=\mathbf{E}_n\begin{bmatrix}
        W_1\\
         \vdots\\
        W_\Ksf
    \end{bmatrix}.
    \label{the encoding matrix}
\end{align}
The communication rate is defined as
the maximum transmission load among all workers, where \begin{align}
\Csf:= \max_{n\in [\Nsf]} \Csf_n. %\frac{|X_n|}{\Lsf}.
\end{align}

Since each worker $n\in[\Nsf]$ only encodes the datasets assigned to it, the coefficients satisfy the encodability constraint $e_{n,i,k}=0$, for any $ i\in[\Csf_n]$ and any $k\in\overline{\Sc_{n}}$. %for each $n\in[\Nsf]$. 
Let $\mathbf{E}=
\begin{bmatrix}
        \mathbf{E}_1\\
         \vdots\\
        \mathbf{E}_\Nsf
    \end{bmatrix}
$ denote the encoding matrix, then the master receives $\mathbf{E}\mathbf{W}$ from the transmissions by the  $\Nsf$ workers.

\text{\textit{2) Decoding Phase:}} After receiving messages from $\Nsf$ workers, the master can apply linear operations to the received messages to recover $\Ksf_\csf$ linear combinations. Accordingly, the decoding process can be expressed as
\begin{align}
    \begin{bmatrix}
        F_1\\
        \vdots\\
        F_{\Ksf_\csf}
\end{bmatrix}=\begin{bmatrix}
        d_{1,1}&d_{1,2}&\ldots&d_{1,\sum_{n\in[\Nsf]}\Csf_n}\\
         d_{2,1}&d_{2,2}&\ldots&d_{2,\sum_{n\in[\Nsf]}\Csf_n}\\
         \vdots& \vdots& \ddots& \vdots\\
          d_{\Ksf_\csf,1}&d_{\Ksf_\csf,2}&\ldots&d_{\Ksf_\csf,\sum_{n\in[\Nsf]}\Csf_n}
    \end{bmatrix}\begin{bmatrix}
        X_1\\
        X_2\\
         \vdots\\
        X_\Nsf
    \end{bmatrix}=\mathbf{D}\mathbf{E}\begin{bmatrix}
        W_1\\
         \vdots\\
        W_\Ksf
    \end{bmatrix},
    \label{the decoding matrix}
\end{align}
where $\mathbf{D}$ is the decoding matrix with dimension $\Ksf_\csf\times\sum_{n\in[\Nsf]}\Csf_n$. From \eqref{the task} and \eqref{the decoding matrix}, we have 
\begin{align}
    \mathbf{F}_1=\mathbf{D}\mathbf{E}.
\end{align}
% The  worst-case   probability of
% error is defined as
% \begin{align}
%  \varepsilon:= \max \Pr\{ \hat{g} \neq g(W_1,   \ldots, W_{\Ksf}) \}. 
% \end{align}
The elements of $\mathbf{F}_1$ are chosen uniformly i.i.d. over $\mathbb{F}_{\asf}$, while the encoding matrix $\mathbf{E}$ is designed to satisfy the encodability constraint.

The probability of error is defined as
\begin{align}
 \varepsilon:= \Pr\{ \mathbf{F}_1 \neq \mathbf{D}\mathbf{E} \}. 
\end{align}

{\bf Objective.} Given any arbitrary heterogeneous data assignment matrix $\mathbf{A}$ satisfying the constraint in~\eqref{eq:data assignment constraint},
if there exists a distributed linear computing scheme with encoding matrices $\{\mathbf{E}_n:n\in [\Nsf]\}$ and a decoding matrix $\mathbf{D}$, such that the communication cost is fixed at $\Csf$, and the master can recover the task function with dimension $\Ksf_\csf$ from the responses of all $\Nsf$ workers  with vanishing error probability $\varepsilon  \stackrel{\Lsf \to \infty}{\longrightarrow}  0$, then the tuple $(\Csf,\Ksf_\csf)$ is said to be achievable. Our objective is to characterize the fundamental tradeoff between the computable dimension 
$\Ksf_\csf$ and the communication cost $\Csf$ under arbitrary heterogeneous data assignment at the workers,  specifically we focus on maximizing the computable dimension under a fixed communication cost.

\subsection{Review of related results}
\label{sub:related}

The following theorem characterizes the minimum communication cost under linear coding by fixing  $\Ksf_\csf=1$.
%In this paper, we go beyond gradient coding, which deals with single demand dimension, to focus on multi-dimensional computational task. This enables us to address more complex distributed  scenarios that involve higher-dimensional computation.
\begin{thm}[\cite{jahani2021optimal}]
\label{thm:result for Kc=1}
For the 
$(\Ksf,\Nsf,\Ksf_\csf=1)$ distributed gradient coding problem with arbitrary heterogeneous data assignment,  the minimum communication cost under linear coding  is 
\begin{align}
\Csf^{\star}=\frac{1}{r},
\label{eq:thm from Kc=1} 
\end{align}
where  $r=\min_{k\in[\Ksf]} |\Cc_k| $ represents the minimum number of workers knowing each dataset. 
\end{thm}

Theorem~\ref{thm:result for Kc=1} shows that for a one-dimensional task, the desired result can be achieved by having the server collect $\mathbf{SFW}$ from the workers' messages. Here, $\mathbf{S}=[
    \mathbf{S}_1;
    \ldots;
    \mathbf{S}_\Nsf]$ is chosen as an MDS matrix, $\mathbf{F} = \begin{bmatrix}
\mathbf{F}_1 \\
\mathbf{F}_2
\end{bmatrix}$, where the virtual demand matrix $\mathbf{F}_2$ is determined according to the encodability constraints, and $\mathbf{W}$ denotes the input vectors. For each column $k \in [\Ksf]$ of $\mathbf{F}$, it is required that $\mathbf{S}_n \mathbf{F}(.,\{k\}) = \boldsymbol{0}$ for all $n \in \overline{\Cc_k}$. Consequently, the dimensionality of the virtual demand is $\Csf \Nsf - \Ksf_\csf$, which must be no less than $\Csf(\Nsf - r)$ in order to satisfy these constraints. However, when $\Ksf_\csf > 1$, the virtual demand may not have sufficient dimensionality to simultaneously satisfy all encodability constraints, which complicates a straightforward extension of the one-dimensional construction to higher-dimensional tasks.

\begin{thm}[\cite{cheng2025novel}]
\label{thm:result for optimal}
For the $(\Ksf,\Nsf,\Ksf_\csf=\sum_{n\in[\Nsf]}\Csf_n)$ heterogeneous distributed system with arbitrary data assignment, each dataset is partitioned into $\Nsf-1$ packets. If for any positive integer $t \in [\Nsf]$,
\begin{equation}
\left| \bigcap_{n \in [t]} \overline{\Sc_n}^{\Nsf-1} \right|
\leq (\Nsf - 1)\!\left( \sum_{n \in [\Nsf]} \Csf_n - \sum_{n \in [t]} \Csf_n \right),
\end{equation}
where
$\overline{\Sc_n}^{\Nsf-1} \triangleq \bigcup_{i\in[0:\Nsf-2]}(\overline{\Sc_n}+i\Ksf)$,
and
$\Csf_n \triangleq 
\frac{\Ksf+(\Nsf-1)|\Sc_n|-\sum_{n^\prime\in[\Nsf]}|\Sc_{n^\prime}|}{\Nsf-1}$,
then there exists a 
$(\Ksf, \Nsf, \Ksf_{\csf} = \sum_{n \in [\Nsf]} \Csf_n)$ coding scheme that achieves the optimal sum communication cost
$\frac{\Nsf \Ksf}{\Nsf - 1}-\sum_{z \in [\Nsf]} \frac{z |\Ic_z|}{\Nsf - 1}$,
where $\Ic_z$ denotes the set of datasets each of which is stored exactly $z$ times.
\end{thm}
After receiving the messages from the workers, the server collects $[\mathbf{S}_1;\ldots;\mathbf{S}_\Nsf]\begin{bmatrix}
\mathbf{F}_1 \\
\mathbf{F}_2
\end{bmatrix}\mathbf{W}$, where the virtual demand matrix $\mathbf{F}_2$ is chosen to satisfy the MDS property. For each $n \in [\Nsf]$, the encoding matrix $\mathbf{S}_n$ can be selected from the left null space of $\mathbf{F}(., \overline{\Sc_n})$ to satisfy the encodability constraints. However, this optimality is restricted to certain data assignment; for arbitrary heterogeneous data assignment, the available encoding degrees of freedom may be insufficient to guarantee the linear independence of the encoding matrices $\{\mathbf{S}_n:n\in[\Nsf]\}$. In our model,
it is for any subset $\Uc \subseteq [\Nsf]$, define $\overline{\Sc_\Uc} = \bigcap_{n \in \Uc} \overline{\Sc_n}$, a necessary condition is
$\Csf \Nsf - \bigl| \overline{\Sc_\Uc} \bigr| \geq \Csf |\Uc|$,
which imposes a fundamental limit on the achievable communication cost.

% \begin{thm}[\cite{cheng2025novel}]
% \label{thm:result for optimal}
% In a $(\Ksf,\Nsf,\Ksf_\csf=\sum_{n\in[\Nsf]}\Csf_n)$ DMTL system, if for any positive integer $t \in [\Nsf]$ it holds that $\left| \bigcap_{n \in [t]} \overline{\Sc_n} \right| \leq \sum_{n\in[\Nsf]}\Csf_n - \sum_{n \in [t]} \Csf_n$, then there exists a $(\Ksf, \Nsf, \Ksf_\csf=\sum_{n\in[\Nsf]}\Csf_n)$ DMTL scheme that achieves the minimum uplink and downlink communication loads.
% \end{thm}

\section{Main Results}
\label{sec:main}
%The recovery of a computational task relies on the transmissions from workers, given that the matrix $\mathbf{A}$ is random due to arbitrary data assignment, the key point of designing the distributed computing scheme is the structure of the matrix $\mathbf{A}$. The sparsity of $\mathbf{A}$ limits the achievable computable dimension.
\subsection{Converse and Achievable Bounds}
 In the distributed linearly separable computation problem with arbitrary heterogeneous data assignment, the problems of maximizing 
$\Ksf_\csf$ under a fixed communication cost $\Csf$, and minimizing 
$\Csf$ under a fixed $\Ksf_\csf$, are equivalent. This is because any point 
($\Csf_1$,$\Ksf_{\csf_1}$) on the capacity region implies that: (1) for a given $\Csf=\Csf_1$, no scheme can achieve a larger computable dimension than $\Ksf_{\csf_1}$; and (2) for a given 
$\Ksf_\csf=\Ksf_{\csf_1}$, no scheme can achieve a smaller communication cost than $\Csf_1$. Thus, these two optimization objectives characterize the same capacity boundary. In this paper, we consider the problem of maximizing 
$\Ksf_\csf$ under a fixed communication cost $\Csf$. 
Obviously, we only need to consider the communication cost $\Csf\in[0, \max_{n\in[\Nsf]} |\Sc_n|]$.

We first define the sparsity property of the matrix $\mathbf{A}$ under a fixed communication cost $\Csf$, which will be used in the proposed converse bound and achievable scheme. Let %we define the set 
   % \begin{align}
    %	\Zc=\{(\Gc,\Qc)&|\Gc\subseteq[\Nsf],\Qc\subseteq[\Ksf],\nonumber \\
    %	&\mathbf{A}(\Gc,\Qc)=\boldsymbol{0}_{|\Gc|\times|\Qc|},|\Gc|+|\Qc|>\Nsf_\rsf \},
   % 	\label{the sparsity of A under C=L}
   % \end{align}
  %  and let $\alpha = \max \left\{ \left| \Gc \right| \,\middle|\, (\Gc, \Qc) \in \Zc \right\}$, $\Gc^\prime=\bigcup\limits_{(\Gc,\Qc)\in\Zc} \Gc$.
    \begin{align}
    	\Zc=\{(\Gc,\Qc)&|\Gc\subseteq[\Nsf],\Qc\subseteq[\Ksf],\nonumber \\
    	&\mathbf{A}(\Gc,\Qc)=\boldsymbol{0}_{|\Gc|\times|\Qc|},\Csf|\Gc|+|\Qc|>\Csf\Nsf \}.
    	\label{the sparsity of A}
    \end{align}
 More intuitively, assume that the submatrix $\mathbf{A}(\Gc, \Qc)$ is all-zero, which implies that none of the workers in $\Gc$ store any datasets from $\Qc$. If $\Csf|\Gc|+|\Qc|>\Csf\Nsf$, the total amount of information the master can receive from the remaining $\Nsf-|\Gc|$ workers is at most $\Csf(\Nsf-|\Gc|)< |\Qc|$, which is insufficient to recover the information corresponding to the datasets in $\Qc$. Thus, the set $\Zc$ identifies the all-zero submatrices in the data assignment matrix, each of which limits communication efficiency in distributed computing.
    %we analyze its structure and identify key characteristic cut sets. for any sub-matrix consisting of \( \alpha_i \) rows from \(\mathbf{A}\), if there exist at most \(\beta_i\) columns where all entries are `0' and  \( \alpha_i + \beta_i > \Nsf_\rsf \), the index set of these \( \alpha_i\) rows  as \( \Gc_i \), and the index set of these \( \beta_i\) columns  as \( \Qc_i \). 

%\begin{thm}
%\label{thm:first result}
%For the $(\Ksf,\Nsf,\Nsf_\rsf,\Csf)$ heterogeneous distributed system with arbitrary data assignment and unit communication cost $\Csf=\Lsf$, any achievable computable dimension should satisfy
%\begin{align}
%\Ksf_\csf\leq\Nsf_\rsf- \alpha.  \label{eq:now first result}
%\end{align}
%\end{thm}

%\begin{thm}
%\label{thm:second result}
%For the $(\Ksf,\Nsf,\Nsf_\rsf,\Csf)$ heterogeneous distributed system with arbitrary data assignment and unit communication cost $\Csf=\Lsf$, the computable dimension is achievable
%\begin{align}
%\Ksf_\csf=\Nsf_\rsf-t .  \label{eq:now second result}
%\end{align}
%\end{thm}
%
%{\bf Extension to Arbitrary Communication Costs.}
%We extend our proposed scheme to accommodate arbitrary communication costs $\Csf=\frac{p}{q}\Lsf$. From \eqref{the sparsity of A under C=L}, we obtain

%For any sub-matrix consisting of \( \alpha_i \) rows from \(\mathbf{A}\), if there exist at most \(\beta_i\) columns where all entries are `0' and  \( p\alpha_i + q\beta_i > p\Nsf_\rsf \), we define the index set of these \( \alpha_i\) rows  as \( \Gc_i \) and \( \Gc = \cup_i \Gc_i \).
\begin{thm}[Converse bound] 
\label{thm:first result}
For the $(\Ksf,\Nsf,\Csf)$ heterogeneous distributed system with arbitrary data assignment and a fixed communication cost  $ \Csf\in[0, \max_{n\in[\Nsf]} |\Sc_n|]$, we let $\alpha = \max \left\{ \left| \Gc \right| \,\middle|\, (\Gc, \Qc) \in \Zc \right\}$, the maximum computable dimension should satisfy
\begin{align}
\Ksf_\csf\leq \min \{\Csf(\Nsf-\alpha),\Ksf\} .  \label{eq:now first result}
\end{align}
\end{thm}
The proof of the above converse bound is provided in Section~\ref{sec:proof of converse}.
The following theorem presents the achievable bound by our proposed scheme.
\begin{thm}[Achievable bound]
\label{thm:second result}
For the $(\Ksf,\Nsf,\Csf)$ heterogeneous distributed system with arbitrary data assignment and a fixed communication cost $\Csf\in[0, \max_{n\in[\Nsf]} |\Sc_n|]$, let $\Gc^\prime=\bigcup\limits_{(\Gc,\Qc)\in\Zc} \Gc$ and \( t \) represent the maximum number of `0' in each column of the sub-matrix $\mathbf{A}(\Gc^\prime,.)$, the computable dimension is achievable
\begin{align}
\Ksf_\csf=\min \{\Csf(\Nsf-t),\Ksf \}.  \label{eq:now second result}
\end{align}
\end{thm}

When $t=\alpha$ in Theorems~\ref{thm:first result} and~\ref{thm:second result}, we can derive the following optimality theorem.
\begin{thm}[Optimality]
	\label{eq:optimality}
	For the $(\Ksf,\Nsf,\Csf)$ heterogeneous distributed system with arbitrary data assignment and a fixed communication cost $\Csf\in[0, \max_{n\in[\Nsf]} |\Sc_n|]$, let $\alpha = \max \left\{ \left| \Gc \right| \,\middle|\, (\Gc, \Qc) \in \Zc \right\}$, when the maximum $|\Gc|$ is equal to the maximum number of `0' in each column of the sub-matrix $\mathbf{A}(\Gc^\prime,.)$, we obtain 
	\begin{align}
		\Ksf_\csf^{\star} = \min \{ \Csf(\Nsf-\alpha),\Ksf \}.  \label{computable capacity}
	\end{align}
\end{thm}

%From Theorem~\ref{eq:optimality}, for arbitrary heterogeneous data assignment, our scheme achieves optimality when the assignment structure satisfies the condition that the maximum $|\Gc|$ is equal to the maximum number of `0' in each column of the sub-matrix $\mathbf{A}(\Gc^\prime,.)$, (i.e., $t=\alpha$).
This result implies that the maximum computable dimension depends on the data assignment structure, while for the case $\Ksf_{\csf}=1$ the optimal communication cost only depends on the minimum number of workers knowing each dataset (as shown in Theorem~\ref{thm:result for Kc=1}).\footnote{\label{foot:offline} For example, consider the storage pattern $\mathbf{A}=\begin{bmatrix}
     0&0 &0 &*&*   \\
 0&0 &0 &*&0   \\
*&*  &*&0&*\\
\end{bmatrix}$. Based on the sparsity characterization in \eqref{the sparsity of A}, we identify the set $\Zc=\{(\{1\},\{1,2,3\}),(\{2\},\{1,2,3,5\}),(\{1,2\},\{1,2,3\})\}$ and $\Gc^\prime=\{1,2\}$ for the case $\Csf=1$. Thus we can obtain the following parameters: the parameter $\alpha=|\{1,2\}|=2$; and $t=2$, representing the maximum number of `0' in each column of the sub-matrix $\mathbf{A}(\{1,2\},.)$. Under these settings, our scheme achieves the optimal computation dimension $	\Ksf_\csf^{\star} =1$. }

\subsection{Comparison with benchmark schemes}

We now compare the performance of our scheme with the benchmark schemes in \cite{jahani2021optimal} and \cite{cheng2025novel}. 

According to \cite{jahani2021optimal}, achieving a one-dimensional computation task requires a minimum communication cost of $\frac{1}{r}$. Thus, to achieve a computation dimension of $\Csf(\Nsf-t)$, where $\Csf$ is the communication cost in our scheme, their required communication cost would be $\frac{\Csf(\Nsf-t)}{r}$. Recall that in our scheme, $\Nsf - r$ represents the maximum number of zeros in each column of the matrix $\mathbf{A}$, and $t$ denotes the maximum number of zeros in each column of the submatrix $\mathbf{A}(\Gc^\prime,.)$ when $\Gc^\prime \neq \emptyset$. We have $\Nsf - r \geq t$, which implies $\frac{\Nsf - t}{r} \geq 1$. When $\Gc^\prime = \emptyset$, we set $t = 0$ to ensure $\frac{\Nsf}{r} \geq 1$. Therefore, our scheme achieves the desired computation dimension with no higher communication cost than required by \cite{jahani2021optimal}.

On the other hand, the scheme proposed in \cite{cheng2025novel} is constructed by choosing the virtual demand matrix $\mathbf{F}_2$ to satisfy the MDS property and selecting $\mathbf{S}_n$ from the left null space of $\mathbf{F}(., \overline{\Sc_n})$ for each $n \in [\Nsf]$ to satisfy the encodability constraints. In our framework, this construction corresponds to the special case where $\Zc = \emptyset$ and $\mathcal{G}^\prime = \emptyset$. Consequently, the communication cost must satisfy
$\Csf \geq \max\left\{ \frac{|\Qc|}{\Nsf - |\Gc|} \Big|\ \Gc\subseteq[\Nsf],\Qc\subseteq[\Ksf],\mathbf{A}(\Gc,\Qc)=\boldsymbol{0}_{|\Gc|\times|\Qc|}\right\}$.

%\textbf{Remark.}
% \begin{rem}
% The key point is that whether we can fully recover the computation depends on how the datasets are distributed across workers. If some datasets are stored by too few workers (i.e., if some columns of the assignment matrix contain too many zeros), then no matter how we process the data, there will not be enough information to recover the full result. Thus, the maximum recoverable dimension is determined by the structure of the assignment matrix, which is the main focus of our analysis.
%  {\blue I CANNOT UNDERSTAND WHAT YOU WANT TO MEAN HERE.} 
% \end{rem}
% \begin{figure}
%   \centering
%   \includegraphics[width=0.43 \textwidth]{picture2.png}
%   \caption{Comparison to the repetition scheme with the data assignment in Example~\ref{ex851}.}
%   \label{fig: dimension comparison}
% \end{figure}

\iffalse
{\bf Consistent with the existing conclusion.}
When we consider the communication cost $\Csf=\frac{1}{r-\Nsf+\Nsf_\rsf}-\epsilon$, where $\epsilon>0$ is an arbitrarily small constant. Since $r$ represents the minimum number `*' in each column in $\mathbf{A}$, it is obvious that there exists $|\Gc|=\Nsf-r$ and $|\Qc|=1$, so that $\Csf\Nsf_\rsf-\Csf|\Gc|=(\frac{1}{r-\Nsf+\Nsf_\rsf}-\epsilon)(\Nsf_\rsf-\Nsf+r)=1-\epsilon(\Nsf_\rsf-\Nsf+r)<|\Qc|$ satisfies the constraint in \eqref{the sparsity of A}, then $t\geq|\Gc|=\Nsf-r$, $\lfloor\Csf(\Nsf_\rsf-t)\rfloor\leq(\Nsf_\rsf-\Nsf+r)(\frac{1}{r-\Nsf+\Nsf_\rsf}-\epsilon)=1-(\Nsf_\rsf-\Nsf+r)\epsilon<1$, $\Ksf_\csf=0$.
\fi
\section{Proof of Theorem~\ref{thm:second result}:Achievable scheme}
\label{sec:proof of subregion}
\subsection{General structure of the scheme}
\label{subsec:the scheme under unit communication cost}
In this section, we focus on the scenario where $\Csf$ is an integer.
After receiving messages from all $\Nsf$ workers, the master can recover $\mathbf{F}\mathbf{W}$, where $\mathbf{F}=\begin{bmatrix}
    \mathbf{F}_1\\
    \mathbf{F}_2
\end{bmatrix}$ has dimension $\Csf\Nsf\times\Ksf$. Here, $ \mathbf{F}_1$ corresponds to the task coefficient matrix with dimension $\Ksf_\csf\times\Ksf$, and 
$ \mathbf{F}_2$ represents the virtual demand coefficient matrix with dimension $(\Csf\Nsf-\Ksf_\csf) \times\Ksf$. 

During the computing phase, each worker is required to transmit a coded message. Specifically, for each worker $n\in[\Nsf]$, the transmitted message is defined as follows:
    \begin{align}
        X_n=\mathbf{S}_n \mathbf{F}\mathbf{W},
        \label{the transmission of worker n}
    \end{align}
    where $\mathbf{S}_n$ is a matrix with dimension $\Csf\times \Csf\Nsf$ and $X_n$ contains $\Csf\Lsf$ symbols. The matrix $\mathbf{S}_n$ will be later determined.
    Each worker transmits a coded message of $\Csf\Lsf$ symbols. Our objective is to find the maximum value of $\Ksf_\csf$, i.e., the maximum linearly independent combinations that can be successfully recovered at the master. To achieve this, we impose constraints on the design of $(\mathbf{S}_n:n\in[\Nsf])$ and $\mathbf{F}_2$. Notably, the first $\Ksf_\csf$ rows of $\mathbf{F}$ (the matrix $\mathbf{F}_1$) are predetermined by the computational task itself, whose elements are i.i.d. over $\mathbb{F}_\asf$. To ensure encodability and decodability, the following two constraints must be satisfied.
    \begin{constraint}[Encodability constraint] 
    \label{pro:encoding} 
    To ensure that each worker only participates in computations related to the datasets it owns, for each worker $n\in[\Nsf]$ we have 
    \begin{align}
\mathbf{S}_n\mathbf{F}(.,\overline{\Sc_{n}})=\boldsymbol{0}_{\Csf\times|\overline{\Sc_{n}}|}.
     \label{encodability}
     \end{align}
    \end{constraint}
    This constraint ensures that in the message $X_n$ sent by worker $n\in[\Nsf]$, the coefficients corresponding to the unassigned datasets in the message are set to `0'.
    \begin{constraint}[Decodability constraint] 
    \label{pro:decoding} 
    To guarantee that the master can recover the desired computation results, the matrix
    \begin{align}
     \begin{bmatrix}
         \mathbf{S}_1\\
         \vdots\\
        \mathbf{S}_\Nsf
     \end{bmatrix} 
    \label{decodability}
     \end{align}
    with dimension $\Csf\Nsf\times\Csf\Nsf$ is full rank.
    \end{constraint}
    After receiving transmissions from $\Nsf$ workers, the master recovers $\mathbf{F}\mathbf{W}$ by computing 
    \begin{align}
    \label{the recovery of F}
        \mathbf{F}\mathbf{W}=\begin{bmatrix}
        \mathbf{S}_1\\
         \vdots\\
        \mathbf{S}_\Nsf
    \end{bmatrix} ^{-1} \begin{bmatrix}
       X_{1} \\
       \vdots \\
       X_{\Nsf}
    \end{bmatrix},
    \end{align}
 where $\begin{bmatrix}
       \mathbf{S}_1\\
         \vdots\\
        \mathbf{S}_\Nsf
    \end{bmatrix}$ is an invertible matrix due to Constraint~\ref{pro:decoding}. Consequently, the master can successfully reconstruct the computational task $\mathbf{F}_1 \mathbf{W}$.
 
\begin{example}[$(\Ksf,\Nsf,\Csf)=(8,5,1)$]
\rm
\label{ex851}
In the following, we introduce one example to illustrate the main idea on the selection of $\{\mathbf{S}_1,\ldots,\mathbf{S}_5\}$ and $\mathbf{F}_2$.
In this example, we consider the data assignment matrix as 
$$\mathbf{A}=\begin{bmatrix}
    0&0 &0 &0&*&*&*&*   \\
 0&0 &0 &0&*&*&*&*   \\
*&*  &*&0&0&0 &0 &0\\
*  &*&*&0&*&0&*&* \\
0&*&*&*&*&*&*&*
\end{bmatrix};$$
if the element at position $(n,k)$ is `0', worker $n$ does not possess dataset $D_k$; otherwise, it does. 

The entries of $\mathbf{A}$ are all `0' in the positions corresponding to the submatrices $\mathbf{A}([2],[4])$ and $\mathbf{A}(\{3\},[4:8])$, satisfying $|\Gc|+|\Qc|>5$.
Thus, $\Zc=\{([2],[4]),(\{3\},[4:8])\}$. We have $\Gc^\prime=\{1,2,3\}$ and $ t=3$;  then obtain $\Ksf_\csf=\min\{\Nsf- t,\Ksf\}=2$. Consider a 2-dimensional task,
\begin{align}
    f(D_1,\ldots, D_8)=\begin{bmatrix}
     F_1\\
     F_2
\end{bmatrix}&=\mathbf{F}_1
\begin{bmatrix}
    W_1\\
    \vdots\\
    W_8
 \end{bmatrix},\end{align}
 the coefficient matrix $\mathbf{F}_1$ is determined by the task, i.e., the elements are selected  uniformly i.i.d. over $\mathbb{F}_{\asf}$.
 
 The master should recover the computational task from the transmissions of $\Nsf=5$ workers,  $$\begin{bmatrix}
    F_1\\
    F_2\\
    F_3\\
    F_4\\
    F_5
\end{bmatrix}=\mathbf{F}\begin{bmatrix}
    W_1\\
    \vdots\\
    W_8
\end{bmatrix}=\begin{bmatrix}
  (\mathbf{F}_1)_{2\times8}\\
  (\mathbf{F}_2)_{3\times8}
 \end{bmatrix}\begin{bmatrix}
 W_1\\
    \vdots\\
    W_8
 \end{bmatrix}.$$
 
{\bf Computing Phase:}
For each worker $n\in[5]$, it can compute  $f_k(D_k)$ and send $W_k$ for each $k\in\Sc_n$. To achieve encodability, we employ the zero-forcing method: 
\begin{itemize}
    \item Generate the encoding matrix of worker $n\in\Gc^\prime.$ We randomly generate $\begin{bmatrix}
        \mathbf{S}_1\\
       \mathbf{S}_2\\
       \mathbf{S}_3
\end{bmatrix}$ with elements uniformly i.i.d. over $\mathbb{F}_{\asf}$, where each encoding matrix contains one vector.
    \item Complete the construction of the matrix $\mathbf{F}_2$. For example, dataset $D_1$ cannot be transmitted by workers $1$ and $2$. Given that $\begin{bmatrix}
        \mathbf{S}_1\\
      \mathbf{S}_2
    \end{bmatrix}\mathbf{F}(.,\{1\})=\begin{bmatrix}
        0\\
        0
    \end{bmatrix}$, where the vectors in $\mathbf{S}_1$ and $\mathbf{S}_2$ are linearly independent, we can determine $\mathbf{F}(.,\{1\})$ as follows:
    select one of its last three elements uniformly i.i.d. over $\mathbb{F}_{\asf}$, and solve the remaining two elements such that the above constraint is satisfied.
   Apply the same procedure to construct each column $\mathbf{F}(.,\{k\})$ for $k\in [2:8]$, using the corresponding constraint vectors. 
%    The resulting matrix is $$\mathbf{F}=\begin{bmatrix}
%   1 & 2 & 5 & 8 & 2 & 2 & 3 & 5 \\
% 1 & 2 & 3 & 4 & 2 & 1 & 7 & 3 \\
% 3 & 3 & 3 & \frac{3}{2} & 2 & 3 & 2 & 2 \\
% 18 & 12 & 4 & -16 & 3 & 3 & 3 & 2 \\
% -34 & -26 & -16 & 15 & -14 & -\frac{31}{2} & -18 & -17
%  \end{bmatrix}.$$
    \item Determine the encoding matrix of worker $n\notin\Gc^\prime.$
    Let us focus on worker $4$, who cannot compute $W_4$ and $W_6$. We retrieve the $i^{\text{th}}$ column of $\mathbf{F}$ where $i\in\overline{\Sc_4}=\{4,6\}$, and obtain one vector in the left null space of ${\mathbf{F}}(.,\{4,6\})$ for worker $4$ as the encoding matrix $\mathbf{S}_4$; similarly we can obtain $\mathbf{S}_5$ for worker $5$.
    
%     Then each worker sends
%    \begin{subequations}
%    \begin{align}
%&X_1=\sv_1\mathbf{F}\mathbf{W}=14W_4;\\
%&X_2=\sv_2\mathbf{F}\mathbf{W}=24W_4;\\
%&X_3=\sv_3\mathbf{F}\mathbf{W}=26W_1+17W_2+35W_3;\\
%&X_4=\sv_4\mathbf{F}\mathbf{W}=W_2-W_3+12W_4;\\
%&X_5=\sv_5\mathbf{F}\mathbf{W}=W_1+2W_2+13W_4.
%  \end{align}
%  \end{subequations}
\end{itemize}

{\bf Decoding Phase:} We can check that the matrix \begin{align}
\label{decoding matrix}
    \begin{bmatrix}
       \mathbf{S}_1\\\mathbf{S}_2\\\mathbf{S}_3\\\mathbf{S}_4\\\mathbf{S}_5
    \end{bmatrix}
%     =\begin{bmatrix}
%           1 & 1 & 4 & 3 & 2 \\
% 2 & 4 & 2 & 5 & 3 \\
% 3 & 1 & 4 & 4 & 2\\-\frac{1}{2} &1 &0 &0 &0\\-3 &0 &1 &0 &0
%     \end{bmatrix}
\end{align} is full rank. Hence, after receiving transmissions from $5$ workers, the master can recover the task function.

The main difficulty in our proposed scheme is to prove that the matrix in \eqref{decoding matrix} is full rank. By our construction, the matrix is obtained through two nested null-space constructions, setting our method apart from most existing works that involve only a single-stage null-space construction\cite{cadambe2009interference,cadambe2008interference,jafar2008degrees}, and making this proof a key technical innovation of our work. For the
sake of simplicity, in the following we provide the sketch of
the feasibility proof.

Recall that for each $n\in\{1,2,3\}$, the matrix $\mathbf{S}_n$ is generated with elements uniformly i.i.d. over $\mathbb{F}_{\asf}$; for each $n\in\{4,5\}$,  the elements of  the matrix $\mathbf{S}_n$ must be determined to satisfy the linear constraints in \eqref{encodability}. Our goal is to prove that the matrix in \eqref{decoding matrix} is full rank with high probability.
The determinant of the matrix  in \eqref{decoding matrix} could be seen as $D= \frac{P}{Q}$, where $P$ and $Q$ are multivariate polynomials whose variables are the elements in  $\{\mathbf{S}_n:n\in\{1,2,3\}\}$, $\mathbf{F}([2]\cup\Fc_k,\{k\})$ for each $k\in[8]$ and $\mathbf{S}_n(.,[5]\setminus\Dc)$ for each $n\in\{4,5\}$, where $\Fc_k$ and $[5]\setminus\Dc$ are index sets whose corresponding elements are chosen uniformly i.i.d. over $\mathbb{F}_{\asf}$. Hence, 
by the  Schwartz--Zippel Lemma~\cite{Schwartz,Zippel,Demillo_Lipton}, if we can further show that the multivariate polynomial  $P$ is non-zero      (i.e., a multivariate polynomial whose coefficients are not all $0$), the probability that this multivariate polynomial is equal to $0$ over all possible realization of the elements in $\{\mathbf{S}_n:n\in\{1,2,3\}\}$, $\mathbf{F}([2]\cup\Fc_k,\{k\})$ for each $k\in[8]$ and $\mathbf{S}_n(.,[5]\setminus\Dc)$ for each $n\in\{4,5\}$,  goes to $0$ when $\asf$ goes to infinity, and thus the matrix in~\eqref{decoding matrix} is full rank with high probability. 
 So in the following, we need to show that $P=DQ$ is   non-zero by finding out a realization of  $\{\mathbf{S}_n:n\in\{1,2,3\}\}$, $\mathbf{F}([2]\cup\Fc_k,\{k\})$ for each $k\in[8]$ and $\mathbf{S}_n(.,[5]\setminus\Dc)$ for each $n\in\{4,5\}$. Now we explicitly construct a realization that guarantees the multivariate polynomial $P$ is non-zero. 
 
 We first generate
\begin{align}
 \begin{bmatrix}
       \mathbf{S}_1\\
       \mathbf{S}_2\\
       \mathbf{S}_3
\end{bmatrix}=
\begin{bmatrix}
0 &0&1&0&0\\
0 &0&0&1&0\\
0 &0&0&0&1
\end{bmatrix},
\end{align} and determine the matrix $\mathbf{F}$ accordingly.  For each column $k\in[\Ksf]$, we randomly select $3-|\overline{\Cc_{k}}\cap\Gc^\prime|$ positions from the last $3$ rows, and assign values to these positions with elements uniformly i.i.d. over $\mathbb{F}_{\asf}$, then solve the remaining $|\overline{\Cc_{k}}\cap\Gc^\prime|$ variables by constraint.\footnote{\label{index detail} The selection of the variable indices, as well as the indices of elements generated uniformly i.i.d. over 
$\mathbb{F}_{\asf}$, can be found in detail in Section \ref{sec:proof of decodability}.} 

For the first column, we have $\overline{\Cc_{1}}\cap\Gc^\prime=\{1,2\}$, so we need to select $1$ position $\Fc_1$ from the last $3$ rows with the element uniformly i.i.d. over $\mathbb{F}_{\asf}$, and solve the remaining $2$ elements by $\begin{bmatrix}
       \mathbf{S}_1\\
       \mathbf{S}_2
    \end{bmatrix}\mathbf{F}(.,\{1\})=\begin{bmatrix}
        0\\
        0
    \end{bmatrix}$. We denote $\mathbf{S}_{k}^\prime$ as the matrix composed of $\{\mathbf{S}_n:n\in\overline{\Cc_{k}}\cap\Gc^\prime\}$, we have
$$\mathbf{S}_{1}^\prime{\mathbf{F}}(.,\{1\})=\begin{bmatrix}
        0\\
        0
    \end{bmatrix},$$ then
\begin{align}
\mathbf{S}_{1}^\prime(.,[3:5]\setminus\Fc_1){\mathbf{F}}([3:5]\setminus\Fc_1,\{1\})=-\mathbf{S}_{1}^\prime(.,\Fc_{1}\cup[2])
{\mathbf{F}}(\Fc_1\cup[2],\{1\}).
\end{align}
By the Cramer's rule, it can be seen that
\begin{align}
   {\mathbf{F}}(\{l\},\{1\})=\frac{\text{det}(\mathbf{Y}_{\overline{1},l})}{\text{det}\Big(\mathbf{S}_{1}^\prime(.,[3:5]\setminus\Fc_1)\Big)}, \forall l\in[3:5]\setminus\Fc_1.
\end{align}
Assuming $l$ is the
$s^{\text{th}}$ smallest value in $[3:5]\setminus\Fc_1$, we define $\mathbf{Y}_{\overline{1},l}$ as the matrix formed by replacing the $s^\text{th}$ column of $\mathbf{S}_{1}^\prime(.,[3:5]\setminus\Fc_1)$ by $-\mathbf{S}_{1}^\prime(.,\Fc_{1}\cup[2])
{\mathbf{F}}(\Fc_1\cup[2],\{1\})$.
To satisfy this constraint, we set $\mathbf{F}([2],\{1\})$ as $\begin{bmatrix}
    *\\
    0
\end{bmatrix},$ where `*' represents a symbol uniformly i.i.d. over $\mathbb{F}_{\asf}$, further select $\Fc_1=\{5\}$ and generate the element in $\mathbf{F}(\Fc_1,\{1\})$  uniformly i.i.d. over $\mathbb{F}_{\asf}$.  It can be seen that $\text{det}\Big(\mathbf{S}_{1}^\prime(.,[3:5]\setminus\Fc_1)\Big)\neq0$, thus the elements $\{\mathbf{F}(\{l\},\{1\}):\forall l\in[3:5]\setminus\Fc_1\}$ exist, then we solve  $ {\mathbf{F}}(\{3\},\{1\})={\mathbf{F}}(\{4\},\{1\})=0$.  Following the same procedure for the remaining columns, we can construct
    \begin{align}
        \mathbf{F}=\begin{bmatrix}
*  &*&*&0&*&0&*&* \\
0&*&*&*&*&*&*&*\\
0&0 &0 &0&*&*&*&*   \\
 0&0 &0 &0&*&*&*&*   \\
*&*  &*&0&0&0 &0 &0
\end{bmatrix},
    \end{align}
with the same form as $\begin{bmatrix}
     \mathbf{A}(\{4\},.)\\
        \mathbf{A}(\{5\},.)\\
       \mathbf{A}(\{1\},.)\\
        \mathbf{A}(\{2\},.)\\
         \mathbf{A}(\{3\},.)
\end{bmatrix}$.

For worker $4$, we retrieve the $i^{\text{th}}$ column of $\mathbf{F}$ where $i\in\overline{\Sc_4}=\{4,6\}$
$$ \mathbf{F}(.,\{4,6\})=\begin{bmatrix}
0&0 \\
*&*\\
0&*   \\
 0&*   \\
0 &0
\end{bmatrix}.$$
By choosing $\Dc=\{2,3\}$, we consider the submatrix of $\mathbf{F}(.,\{4,6\})$ formed by selecting rows with indices in $\mathcal{D}=\{2,3\}$, the elements located at the right-diagonal of $\mathbf{F}(\Dc,\{4,6\})$ are all `*', where `*' represents a symbol uniformly i.i.d. over $\mathbb{F}_{\asf}$. Thus, the matrix $ \mathbf{F}(.,\{4,6\})$ has rank equal to $2$ and hence is of column-wise full rank.

We now construct $\mathbf{S}_4$ as follows: set the elements of $\mathbf{S}_4(.,[5]\setminus\Dc)$  uniformly i.i.d. over $\mathbb{F}_{\asf}$, and let $\mathbf{S}_4(.,\{1,4,5\})=[1,0,0]$. Then, we have 
\begin{align}
\mathbf{S}_4(.,\Dc) \mathbf{F}(\Dc,\{4,6\})=-\mathbf{S}_4(.,[5]\setminus\Dc)
\mathbf{F}([5]\setminus\Dc,\{4,6\})=\boldsymbol{0}.
\end{align}
By the Cramer's rule, it can be seen that
\begin{align}
   \mathbf{S}_4(.,\{l\})=\frac{\text{det}(\mathbf{Y}_{\overline{4},l}^\prime)}{\text{det}\Big({\mathbf{F}}(\Dc,\{4,6\})\Big)}, \forall l\in\Dc,
\end{align}
Assuming $l$ is the
$s^{\text{th}}$ smallest value in $\Dc$, we define $\mathbf{Y}_{\overline{4},l}^\prime$ as the matrix formed by replacing the $s^\text{th}$ row of ${\mathbf{F}}(\Dc,\{4,6\})$ by $-\mathbf{S}_4(.,[5]\setminus\Dc)
{\mathbf{F}}([5]\setminus\Dc,\{4,6\})$, 
thereby enabling us to determine $\mathbf{S}_4(.,\{2,3\})=[0,0]$. Thus, we can obtain $\mathbf{S}_4=[1,0,0,0,0]$; similarly, for worker $5$ we have 
$\mathbf{S}_5=[0,1,0,0,0]$.

Hence, we have 
\begin{align}
    \begin{bmatrix}
       \mathbf{S}_1\\\mathbf{S}_2\\\mathbf{S}_3\\\mathbf{S}_4\\\mathbf{S}_5
    \end{bmatrix}
    =\begin{bmatrix}
          1 & 0 & 0 & 0 & 0 \\
0 & 1 & 0 & 0& 0 \\
0 & 0 & 1 & 0 & 0\\0 &0 &0 &1 &0\\0 &0 &0 &0 &1
    \end{bmatrix},
\end{align}
the matrix in \eqref{decoding matrix} is an identity matrix and is thus full rank.

We are going to summarize the main idea in this proposed scheme for integer communication costs, the key points are as follows:
\begin{itemize}
	\item First, for the data assignment matrix \(\mathbf{A}\), we define 
	\begin{align}
		\Zc=\{(\Gc,\Qc)&|\Gc\subseteq[\Nsf],\Qc\subseteq[\Ksf],\nonumber \\
		&\mathbf{A}(\Gc,\Qc)=\boldsymbol{0}_{|\Gc|\times|\Qc|}, \Csf|\Gc|+|\Qc|>\Csf\Nsf \},\label{define set}
	\end{align}
	and let $\alpha = \max \left\{ \left| \Gc \right| \,\middle|\, (\Gc, \Qc) \in \Zc \right\}$, thus we have the converse $\Ksf_\csf\leq \min \{\Csf(\Nsf-\alpha),\Ksf\}$. 
	%    we study its column structure to find the maximum number of \(\beta_i\) columns in which all entries are `0', satisfying the condition \( \alpha_i + \beta_i > \Nsf_\rsf \). Then we define the index set of these rows as  \( \Gc_i \).
	\item Then we define $\Gc^\prime=\bigcup\limits_{(\Gc,\Qc)\in\Zc} \Gc$ and consider the sub-matrix $\mathbf{A}(\Gc^\prime,.)$. \( t \) represents the maximum number of `0' in each column of the sub-matrix $\mathbf{A}(\Gc^\prime,.)$. We generate a $\Csf|\Gc^\prime|\times\Csf\Nsf$ matrix with elements uniformly i.i.d. over $\mathbb{F}_{\asf}$  as $[\mathbf{S}_{\Gc^\prime(1)};\mathbf{S}_{\Gc^\prime(2)};\ldots;\mathbf{S}_{\Gc^\prime(|\Gc^\prime|)}]$. For each $k\in[\Ksf]$, the first $\Ksf_\csf$ elements in ${\mathbf{F}}(.,\{k\})$ are predetermined, and the remaining elements should satisfy 
	\begin{align}
		\mathbf{S}_n{\mathbf{F}}(.,\{k\})=\boldsymbol{0}_{\Csf\times1}, \forall n\in \overline{\Cc_{k}}\cap\Gc^\prime,
		\label{computation constraint}
	\end{align}
	where $\overline{\Cc_{k}}\cap\Gc^\prime$ represents the set of workers in $\Gc^\prime$ that do not store dataset $D_k$ and cannot compute $W_k$. The number of such workers, $|\overline{\Cc_{k}}\cap\Gc^\prime|$, corresponds to the number of `0' in  $\mathbf{A}(\Gc^\prime,\{k\})$. Since $|\overline{\Cc_{k}}\cap\Gc^\prime|\leq t$, we randomly select $\Csf(t-|\overline{\Cc_{k}}\cap\Gc^\prime|)$ positions from the last $\Csf t$ rows, and assign values to these positions with elements uniformly i.i.d. over $\mathbb{F}_{\asf}$. Then we solve the remaining $\Csf|\overline{\Cc_{k}}\cap\Gc^\prime|$ elements in ${\mathbf{F}}(.,\{k\})$ by \eqref{computation constraint}. This process allows us to fully construct the matrix $\mathbf{F}$.
 \item Finally, for each worker $n\in[\Nsf]\setminus\Gc^\prime$, we form $\mathbf{S}_n$ by selecting $\Csf$  linearly independent vectors from the left null space of ${\mathbf{F}}(.,\overline{\Sc_{n}})$. This selection is feasible because for each $n\in[\Nsf]\setminus\Gc^\prime$, row $n$ of $\mathbf{A}$ satisfies $\Csf|\Gc|+|\Qc|\leq \Csf\Nsf$ when $|\Gc|=1$, we have 
    \begin{align}
|\Qc|=|\overline{\Sc_{n}}|\leq \Csf(\Nsf-1),
\label{the sparsity}
    \end{align}
   based on this inequality, we have
    \begin{align}
   \Csf\Nsf-|\overline{\Sc_{n}}|
     \geq \Csf\Nsf-\Csf(\Nsf-1)=\Csf.
     \label{existence}
       \end{align} 
\end{itemize}
Following the construction of the matrix 
$\mathbf{F}$, the last $\Csf t$ rows are partially determined by solving the linear equations imposed by the constraint in~\eqref{computation constraint}. As a result, only the first $\Csf(\Nsf-t)$ rows of 
$\mathbf{F}$ can be freely chosen, with elements uniformly i.i.d. over $\mathbb{F}_{\asf}$.
Therefore, we conclude that the scheme achieves the computable dimension $\Ksf_\csf=\min \{\Csf(\Nsf-t),\Ksf\}$.
% We demonstrate that the proposed scheme satisfies Constraints~\ref{pro:encoding}-\ref{pro:decoding}.
\begin{figure} 
  \centering
    \centering
    \includegraphics[scale=0.55]{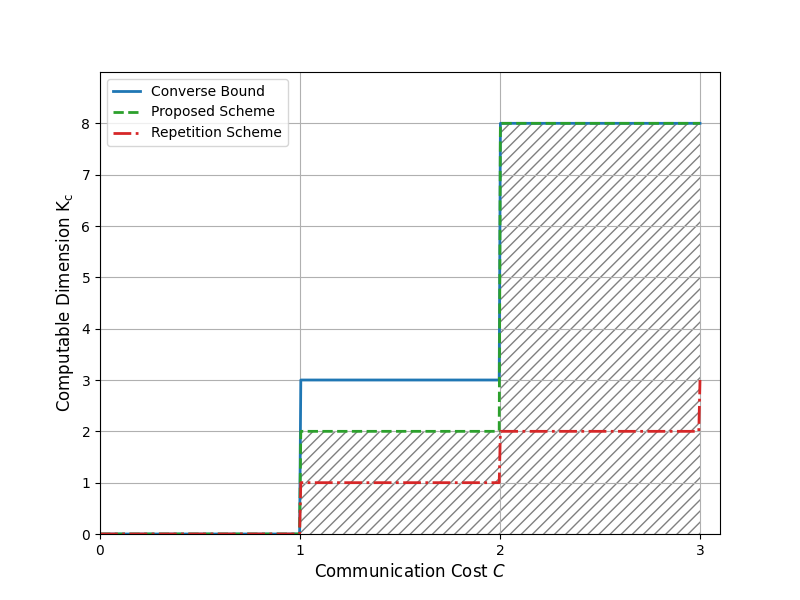}
  \caption{Comparison with the data assignment matrix in Example~\ref{ex851} for integer communication costs.}
  \label{fig:myfig}
\end{figure}

{\bf Constraint~\ref{pro:encoding}.} For each worker $n\in\Gc^\prime$, the encodability is directly ensured by \eqref{computation constraint}; and for each worker $n\notin\Gc^\prime$, the encodability is guaranteed by selecting vectors from the left null space of ${\mathbf{F}}(.,\overline{\Sc_{n}})$ as $\mathbf{S}_n$. 

To address decodability, we present the following lemma, with its detailed proof provided in Section~\ref{sec:proof of decodability}.
\begin{lem}
	\label{lem:full rank of sk}
	With this design, the matrix  $\begin{bmatrix}
\mathbf{S}_1\\
\mathbf{S}_2\\
\vdots\\
\mathbf{S}_{\Nsf}\end{bmatrix}$ is full rank with high probability.
\end{lem}

{\bf Constraint~\ref{pro:decoding}.} From $\Csf\Nsf$  linear combinations of $F_1,\ldots,F_{\Csf\Nsf}$, the decodability constraint can be proven directly by Lemma~\ref{lem:full rank of sk}.

With the data assignment matrix in Example~\ref{ex851}, the tradeoff between the communication cost and the computable dimension for the converse bound, the proposed scheme, and the benchmark scheme which repeats the scheme in~\cite{jahani2021optimal} $\Ksf_{\rm c}$ times (referred to as  Repetition Scheme) is presented in Fig.~\ref{fig:myfig}. The shaded region on the right-hand side of the curve for the proposed scheme denotes its achievable region. As shown in the figure, the proposed scheme achieves a higher computable dimension than the benchmark scheme under the same communication cost.

% Due to the limitation of pages, the detailed structure of the proposed scheme is provided in Appendix~\ref{sec:general structure for C=L}.
%Assume $\Ac=\{1,2,3,4\}$, the master can receive
%\begin{align}
%\begin{bmatrix}
%    X_1\\
%    X_2\\
%    X_3\\
%    X_4
%\end{bmatrix}=\begin{bmatrix}
%    2&1&1&1\\
%    3&2&1&2\\
%    3&5&-1&-4\\
%    4&0&1&1
%\end{bmatrix}
%\mathbf{F}\mathbf{W};
%\end{align} then by computing
%\begin{align}
%\mathbf{F}\mathbf{W}=\begin{bmatrix}
%    2&1&1&1\\
%    3&2&1&2\\
%    3&5&-1&-4\\
%    4&0&1&1
%\end{bmatrix}^{-1}\begin{bmatrix}
%    X_1\\
%    X_2\\
%    X_3\\
%    X_4
%\end{bmatrix},
%\end{align}
%the master can recover the task function $\mathbf{F}_1\mathbf{W}$ from the first two rows of the recovered result.

%Consider the matrix
%$$\begin{bmatrix}
%    \sv_1\\
%    \sv_2\\
%    \sv_3\\
%    \sv_4\\
%    \sv_5
%\end{bmatrix}=\begin{bmatrix}
%    2&1&1&1\\
%    3&2&1&2\\
%    3&5&-1&-4\\
%    4&0&1&1\\
%    5&0&1&1
%\end{bmatrix},$$ 
\hfill $\square$ 
\end{example}
\subsection{Proof of Lemma~\ref{lem:full rank of sk}}
\label{sec:proof of decodability}
In the following, we give the proof of Lemma~\ref{lem:full rank of sk} for the decodability.

Each worker $n\in[\Nsf]$ sends
\begin{align}
    \mathbf{S}_{n}\mathbf{F}
    \begin{bmatrix}
        W_1 \\
        \vdots \\
        W_\Ksf
    \end{bmatrix},
\end{align}
where $\mathbf{S}_{n}\mathbf{F}(.,\overline{\Sc_{n}})=\boldsymbol{0}_{\Csf\times|\overline{\Sc_{n}}|}$. 
We define the set $[\Nsf]\setminus\Gc^\prime$ as $\overline{\Gc^\prime}$. For each $n\in\Gc^\prime$, $\mathbf{S}_{n}$ is generated with elements uniformly i.i.d. over $\mathbb{F}_{\asf}$, so next we discuss the existence of $\{\mathbf{S}_{n}:n\in\overline{\Gc^\prime}\}$.
For each $n\in\overline{\Gc^\prime}$, the matrix $\mathbf{F}(.,\overline{\Sc_{n}})$ has dimension of $\Csf\Nsf\times|\overline{\Sc_{n}}|$, whose rank is $|\overline{\Sc_{n}}|$ with high probability and $|\overline{\Sc_{n}}|\leq\Csf(\Nsf-1)$ (all-zero matrix of $\mathbf{F}(.,\overline{\Sc_{n}})$ satisfies $\Csf|\Gc|+|\Qc|\leq\Csf\Nsf$). We uniformly at random select $\Csf$ vectors from the left null space of $\mathbf{F}(.,\overline{\Sc_{n}})$ as $\mathbf{S}_{n}$; equivalently, to generate $\mathbf{S}_{n}$   we can first select  $\Csf\Nsf-|\overline{\Sc_{n}}|$ column indices in $\mathbf{S}_n$, then set the elements at these columns to be uniformly i.i.d. over $\mathbb{F}_{\asf}$, and finally determine the elements of remaining $|\overline{\Sc_{n}}|$ columns by solving the constraint $\mathbf{S}_n\mathbf{F}(.,\overline{\Sc_{n}})=\boldsymbol{0}_{\Csf\times|\overline{\Sc_{n}}|}$.

Recall that for each $k\in\overline{\Sc_{n}}$, the first $\Ksf_\csf=\Csf(\Nsf-t)$ elements are determined by the task, we randomly select $\Csf(t-|\overline{\Cc_{k}}\cap\Gc^\prime|)$ positions from the last $\Csf t$ rows, and assign values to these positions with elements uniformly i.i.d. over $\mathbb{F}_{\asf}$, we denote these row indices as $\Fc_k$. Then we solve the remaining $\Csf(|\overline{\Cc_{k}}\cap\Gc^\prime|)$ elements in ${\mathbf{F}}(.,\{k\})$ by \eqref{computation constraint}. 
We denote $\mathbf{S}_{k}^\prime$ as the matrix composed of  $\{\mathbf{S}_n:n\in\overline{\Cc_{k}}\cap\Gc^\prime\}$ 
with dimension $\Csf|\overline{\Cc_{k}}\cap\Gc^\prime|\times\Csf\Nsf$, we have
$$\mathbf{S}_{k}^\prime{\mathbf{F}}(.,\{k\})=[0,\ldots,0]^\Tsf,$$ then
\begin{align}
&\mathbf{S}_{k}^\prime(.,[\Csf(\Nsf-t)+1:\Csf\Nsf]\setminus\Fc_k){\mathbf{F}}([\Csf(\Nsf-t)+1:\Csf\Nsf]\setminus\Fc_k,\{k\})\nonumber\\&=-\mathbf{S}_{k}^\prime(.,\Fc_{k}\cup[\Csf(\Nsf-t)])
{\mathbf{F}}(\Fc_k\cup[\Csf(\Nsf-t)],\{k\}).
\label{matrix multiplication}
\end{align}
By the Cramer's rule, it can be seen that
\begin{align}
   {\mathbf{F}}(\{l\},\{k\})=\frac{\text{det}(\mathbf{Y}_{\overline{k},l})}{\text{det}\Big(\mathbf{S}_{k}^\prime(.,[\Csf(\Nsf-t)+1:\Csf\Nsf]\setminus\Fc_k)\Big)}, \forall l\in[\Csf(\Nsf-t)+1:\Csf\Nsf]\setminus\Fc_k,
    \label{solution for the Cramer's rule}
\end{align}
Assuming $l$ is the
$s^{\text{th}}$ smallest value in $[\Csf(\Nsf-t)+1:\Csf\Nsf]\setminus\Fc_k$, we define $\mathbf{Y}_{\overline{k},l}$ as the matrix formed by replacing the $s^\text{th}$ column of $\mathbf{S}_{k}^\prime(.,[\Csf(\Nsf-t)+1:\Csf\Nsf]\setminus\Fc_k)$ by $-\mathbf{S}_{k}^\prime(.,\Fc_{k}\cup[\Csf(\Nsf-t)])
{\mathbf{F}}(\Fc_k\cup[\Csf(\Nsf-t)],\{k\})$. 

It can be seen that $\text{det}\Big(\mathbf{S}_{k}^\prime(.,[\Csf(\Nsf-t)+1:\Csf\Nsf]\setminus\Fc_k)\Big)$
is the determinant of a $\Csf|\overline{\Cc_{k}}\cap\Gc^\prime|\times\Csf|\overline{\Cc_{k}}\cap\Gc^\prime|$ matrix, which can be viewed as a multivariate polynomial whose variables are the elements in $\mathbf{S}_n$ where $n\in\Gc^\prime$. Since the elements in $\{\mathbf{S}_n:n\in\Gc^\prime\}$ are uniformly i.i.d. over $\mathbb{F}_{\asf}$, there is a high probability that the multivariate polynomial $\text{det}\Big(\mathbf{S}_{k}^\prime(.,[\Csf(\Nsf-t)+1:\Csf\Nsf]\setminus\Fc_k)\Big)$ is a non-zero multivariate polynomial with degree $\Csf|\overline{\Cc_{k}}\cap\Gc^\prime|$. Hence, by the Schwartz--Zippel Lemma~\cite{Schwartz,Zippel,Demillo_Lipton}, we have
\begin{subequations}
\begin{align}
\Pr\{ {\mathbf{F}}(\{l\},\{k\}) \text{ exists} \}
&= \Pr \left\{ \text{det}\Big(\mathbf{S}_{k}^\prime(.,[\Csf(\Nsf-t)+1:\Csf\Nsf]\setminus\Fc_k)\Big) \text{ is non-zero} \right\}\\
&\geq 1-\frac{\Csf|\overline{\Cc_{k}}\cap\Gc^\prime|}{\asf}.\label{the pro of existence}
\end{align}
\end{subequations}
%Note that the probability in \eqref{the pro of existence} is taken over all possible realizations of \( \{ \sv_k : k \in [\Ksf] \} \) whose elements are uniformly i.i.d. over $\mathbb{F}_{\asf}$. 
 By the probability union bound, we have
\begin{align}
\Pr \left\{   {\mathbf{F}}(\{l\},\{k\}) \text{ exists}, \forall l\in[\Csf(\Nsf-t)+1:\Csf\Nsf]\setminus\Fc_k, k\in\overline{\Sc_{n}} \right\}
\geq 1 - \frac{|\overline{\Sc_{n}}|\Csf^2|\overline{\Cc_{k}}\cap\Gc^\prime|^2}{\asf}
\xrightarrow{\asf \to \infty} 1.
\label{all pro of existence}
\end{align}
Therefore, we prove that $\mathbf{F}(.,\overline{\Sc_{n}})$ for each $n\in\overline{\Gc^\prime}$ exists with high probability.

Next, we will prove that the matrix $\mathbf{F}(.,\overline{\Sc_{n}})$ is column-full rank with high probability. By selecting rows with indices in $\Dc$, where $|\Dc|=|\overline{\Sc_{n}}|$ and $\Dc\subseteq[\Csf\Nsf] $, if a full-rank sub-matrix of dimension $|\overline{\Sc_{n}}|\times|\overline{\Sc_{n}}|$ can be obtained, then
the matrix $\mathbf{F}(.,\overline{\Sc_{n}})$ is column-full rank.
We expand the determinant of \(\mathbf{F}(\Dc,\overline{\Sc_{n}}) \) as follows:
\begin{align}
\det\left(\mathbf{F}(\Dc,\overline{\Sc_{n}})\right) = \sum_{i \in \left[\left(|\overline{\Sc_{n}}|\right)!\right]} \frac{P_i}{Q_i},
\end{align}
which contains \( \left(|\overline{\Sc_{n}}|\right)! \) terms.  Each term can be expressed as \( \frac{P_i}{Q_i} \), where \( P_i \) and \( Q_i \) are multivariate polynomials whose variables are the elements in $\mathbf{S}_n$ where  $n\in\Gc^\prime$ and the elements in ${\mathbf{F}}(\Fc_k\cup[\Csf(\Nsf-t)],\{k\})$ where  $k\in\overline{\Sc_{n}}$. From \eqref{solution for the Cramer's rule}, it can be seen that each element in \( \mathbf{F}(\Dc,\overline{\Sc_{n}}) \) has degree no more than \( 2\Csf|\overline{\Cc_{k}}\cap\Gc^\prime| \), whose variables are the elements in $\mathbf{S}_n$ where  $n\in\Gc^\prime$ and the elements in ${\mathbf{F}}(\Fc_k\cup[\Csf(\Nsf-t)],\{k\})$ where  $k\in\overline{\Sc_{n}}$. In addition, each term in \( \det\left(\mathbf{F}(\Dc,\overline{\Sc_{n}}) \right)\) is a multivariate polynomial whose variables are the elements in \(\mathbf{F}(\Dc,\overline{\Sc_{n}}) \) with degree \( |\overline{\Sc_{n}}| \). Hence, \( P_i \) and \( Q_i \) are multivariate polynomials  with degree no more than \( 2\Csf|\overline{\Cc_{k}}\cap\Gc^\prime||\overline{\Sc_{n}}| \),  whose variables are the elements in  $\mathbf{S}_n$ where  $n\in\Gc^\prime$ and the elements in ${\mathbf{F}}(\Fc_k\cup[\Csf(\Nsf-t)],\{k\})$ where  $k\in\overline{\Sc_{n}}$.

We then let
\begin{align}
P_{n}^\prime := \det\left(\mathbf{F}(\Dc,\overline{\Sc_{n}})\right) \prod_{i \in \left[\left(|\overline{\Sc_{n}}|\right)!\right]} Q_{i}.
\end{align}

If $\prod_{i \in \left[\left(|\overline{\Sc_{n}}|\right)!\right]} Q_{i} \neq 0$  (i.e., $\mathbf{F}(\Dc,\overline{\Sc_{n}})$ exists) and $P_{n}^\prime \neq 0$, we have $\det\left(\mathbf{F}(\Dc,\overline{\Sc_{n}})\right) \neq 0$ and thus $\mathbf{F}(\Dc,\overline{\Sc_{n}})$ is full-rank.  
To apply the   Schwartz--Zippel Lemma~\cite{Schwartz,Zippel,Demillo_Lipton} in proving that $P_n^\prime$ is non-zero with high probability,  we need to guarantee that  $P_{n}^\prime$ is a non-zero  multivariate polynomial by finding out a realization  such that $P_{n}^\prime \neq 0$ (i.e., $\prod_{i \in \left[\left(|\overline{\Sc_{n}}|\right)!\right]} Q_{i}\neq 0 $ and $\text{det}\left(\mathbf{F}(\Dc,\overline{\Sc_{n}})\right) \neq 0$  at the same time), this specific realization will be proved later.

Then
 by the Schwartz--Zippel Lemma~\cite{Schwartz,Zippel,Demillo_Lipton}, we have
\begin{align}
\Pr \{P_{n}^\prime \neq 0\} \geq 1 - \frac{2\left(|\overline{\Sc_{n}}|\right)! \Csf|\overline{\Cc_{k}}\cap\Gc^\prime||\overline{\Sc_{n}}|}{\asf}. \label{Pknoteq0}
\end{align}

Thus, we have
\begin{align}
\Pr \{\mathbf{F}(.,\overline{\Sc_{n}})\text{ is column-full rank}\}  \geq 1 - \frac{2\binom{\Csf\Nsf}{|\overline{\Sc_{n}}|}\left(|\overline{\Sc_{n}}|\right)!  \Csf|\overline{\Cc_{k}}\cap\Gc^\prime||\overline{\Sc_{n}}|}{\asf}. \label{Pknoteq01}
\end{align}

Hence, from \eqref{all pro of existence} and \eqref{Pknoteq01},  for each $n\in\overline{\Gc^\prime}$ we have
\begin{subequations}
    \begin{align}
&\Pr \{\mathbf{S}_n \text{ exists}, \forall n\in\overline{\Gc^\prime}\} \\
&\geq 1 - \sum_{n\in\overline{\Gc^\prime} }\left(\Pr \{\mathbf{F}(.,\overline{\Sc_{n}})\text{ does not exist}\} + \Pr \{\mathbf{F}(.,\overline{\Sc_{n}})\text{ is not column-full rank}\} \right) \\
&\geq 1 -|\overline{\Gc^\prime}|\left( \frac{|\overline{\Sc_{n}}|\Csf^2|\overline{\Cc_{k}}\cap\Gc^\prime|^2}{\asf}+\frac{2\binom{\Csf\Nsf}{|\overline{\Sc_{n}}|}\left(|\overline{\Sc_{n}}|\right)! \Csf|\overline{\Cc_{k}}\cap\Gc^\prime||\overline{\Sc_{n}}|}{\asf}\right)\xrightarrow{\asf \to \infty} 1.
\end{align}
\end{subequations}

This guarantees the 
the existence of $\mathbf{S}_{n}$ for each $n\in\overline{\Gc^\prime}$. We have
$$\mathbf{S}_{n}{\mathbf{F}}(.,\overline{\Sc_{n}})=[0,\ldots,0],$$ let the elements at $\mathbf{S}_n(.,[\Csf\Nsf]\setminus\Dc)$ for each $n\in\overline{\Gc^\prime}$ uniformly i.i.d. over $\mathbb{F}_{\asf}$, then for each $n\in\overline{\Gc^\prime}$ 
\begin{align}
\mathbf{S}_n(.,\Dc){\mathbf{F}}(\Dc,\overline{\Sc_{n}})=-\mathbf{S}_n(.,[\Csf\Nsf]\setminus\Dc)
{\mathbf{F}}([\Csf\Nsf]\setminus\Dc,\overline{\Sc_{n}}).
\label{matrix multiplication2}
\end{align}
By the Cramer's rule, it can be seen that
\begin{align}
  \mathbf{S}_n(\{l_1\},\{l_2\})=\frac{\text{det}(\mathbf{Y}_{\overline{n},l_1,l_2}^\prime)}{\text{det}\Big({\mathbf{F}}(\Dc,\overline{\Sc_{n}})\Big)}, \forall l_1\in[\Csf] ,l_2\in\Dc,
    \label{solution for the Cramer's rule2}
\end{align}
Assuming $l_2$ is the
$s^{\text{th}}$ smallest value in $\Dc$, we define $\mathbf{Y}_{\overline{n},l_1,l_2}^\prime$ as the matrix formed by replacing the $s^\text{th}$ row of ${\mathbf{F}}(\Dc,\overline{\Sc_{n}})$ by $-\mathbf{S}_n(\{l_1\},[\Csf\Nsf]\setminus\Dc)
{\mathbf{F}}([\Csf\Nsf]\setminus\Dc,\overline{\Sc_{n}})$. 

In the following, we will prove that matrix
\begin{align}
\begin{bmatrix}
    \mathbf{S}_{1}\\
     \mathbf{S}_{2}\\
    \vdots\\
    \mathbf{S}_{\Nsf}
\end{bmatrix}
\label{the matrix of sk}
\end{align}
is full-rank with high probability. We denote the matrix in \eqref{the matrix of sk} by $\mathbf{T}$. We expand the determinant of \( \mathbf{T} \) as follows:
\begin{align}
\det(\mathbf{T}) = \sum_{i \in \left[(\Csf\Nsf)!\right]} \frac{T_i}{K_i},
\end{align}
which contains \( \left(\Csf\Nsf\right)! \) terms.  Each term can be expressed as \( \frac{T_i}{K_i} \), where \( T_i \) and \( K_i \) are multivariate polynomials whose variables are the elements in  $\{\mathbf{S}_{n}:n\in\Gc^\prime\}$ and the randomly selected elements in ${\mathbf{F}}(\Fc_k\cup[\Csf(\Nsf-t)],\{k\})$ for each $k\in[\Ksf]$ and $\mathbf{S}_n(.,[\Csf\Nsf]\setminus\Dc)$ for each $n\in\overline{\Gc^\prime}$. 

We then let
\begin{align}
T^\prime := \det\left(\mathbf{T}\right) \prod_{i \in \left[(\Csf\Nsf)!\right]} K_{i},
\end{align} if $\{\mathbf{S}_{n}:n\in[\Nsf]\}$ exists ($\prod_{i \in \left[(\Csf\Nsf)!\right]} K_{i}\neq0$) and $T^\prime\neq0$, we have $\det\left(\mathbf{T}\right) \neq0$ and thus the matrix in \eqref{the matrix of sk} is full rank. To apply the Schwartz--Zippel Lemma~\cite{Schwartz,Zippel,Demillo_Lipton}, we need to
guarantee that $T^\prime$ is a non-zero multivariate polynomial. To
this end, we only need one specific realization of $\mathbf{F}$ and $\{\mathbf{S}_{n}:n\in[\Nsf]\}$ so that
$T^\prime\neq0$ (or alternatively $\text{det}(\mathbf{T})\neq0$ and $\prod_{i \in \left[(\Csf\Nsf)!\right]} K_{i}\neq0$ at the same time).

So in the following,  we will show that $T^\prime$ and $P_{n}^\prime$ are non-zero multivariate polynomials by finding out a realization.

Recall that in each column of $\mathbf{F}$, we randomly select $\Csf( t- |\overline{\Cc_{k}} \cap \Gc^\prime|)$ positions from the last $\Csf t$ rows, and assign values to these positions with elements uniformly i.i.d. over $\mathbb{F}_{\asf}$. The remaining $\Csf|\overline{\Cc_{k}} \cap \Gc^\prime|$ entries are then determined by the encodability constraint in~\eqref{computation constraint}. According to Theorem~\ref{thm:second result}, we have $\Ksf_\csf = \min\{\Csf(\Nsf - t), \Ksf\}$. Since $\Csf\Nsf - \Ksf_\csf \geq \Csf t$, we can always select $\Csf|\overline{\Cc_{k}} \cap \Gc^\prime|$  variable entries from the last $\Csf t$ rows. For each column $k\in[\Ksf]$, let $\Fc_k$ and $\Vc_k$ denote the sets of row indices corresponding to the elements drawn uniformly i.i.d. over $\mathbb{F}_{\asf}$ and variable entries, respectively.  In the following, we need to show that:
\begin{enumerate}
\item $\text{def}\left(\mathbf{S}_k^\prime(.,\Vc_k)\right)\neq0$ for each $k\in[\Ksf]$, such that ${\mathbf{F}}(.,\{k\})$ exists.
\item $\mathbf{F}(.,\overline{\Sc_{\overline{\Gc^\prime}(i)}})$ for each $i\in[|\overline{\Gc^\prime}|]$ is column-wise full rank, such that $\mathbf{S}_{\overline{\Gc^\prime}(i)}$ exists.
\item $\text{def}\left(\mathbf{T}\right)\neq0$.
\end{enumerate}

Recall $\Gc^\prime=\bigcup\limits_{(\Gc,\Qc)\in\Zc} \Gc$ and \( t \) represents the maximum number of `0' in each column of the sub-matrix $\mathbf{A}(\Gc^\prime,.)$. We will prove in two cases (i) $|\Gc^\prime|=t$ and (ii) $|\Gc^\prime|>t$.

(i) $|\Gc^\prime|=t$.
We first determine the choice of \( \{\mathbf{S}_n: n \in \Gc^\prime\} \). For each $i\in[|\Gc^\prime|]$, we can select \(\mathbf{S}_{\Gc^\prime(i)} \) to be \begin{align}
    \begin{bmatrix}
0 & \cdots & 0  & 1 & 0 & \cdots &0  & 0 & \cdots & 0\\
0  & \cdots & 0 & 0 & 1 & \cdots & 0 & 0 & \cdots & 0\\
\vdots  & \ddots & \vdots & \vdots & \vdots & \ddots & \vdots & \vdots  & \ddots & \vdots\\
0  & \cdots & 0 &  0 & 0 & \cdots & 1  & 0 & \cdots & 0
    \end{bmatrix}
    \label{SG design},
\end{align}
where we have $$\mathbf{S}_{\Gc^\prime(i)}(.,[\Csf(\Nsf-|\Gc^\prime|+i-1)+1:\Csf(\Nsf-|\Gc^\prime|+i)])=\mathbf{I}_{\Csf\times \Csf}. $$
Thus we have 
\begin{align}
 \begin{bmatrix}
       \mathbf{S}_{\Gc^\prime(1)}\\
      \mathbf{S}_{\Gc^\prime(2)}\\
        \ldots\\
\mathbf{S}_{\Gc^\prime(|\Gc^\prime|)}
    \end{bmatrix}=
\left[
\begin{array}{c@{\hskip 6pt}c@{\hskip 6pt}c}
\boldsymbol{0}_{\Csf|\Gc^\prime|\times\Csf(\Nsf-|\Gc^\prime|)} & \vdashline&  \mathbf{I}_{\Csf|\Gc^\prime|\times\Csf|\Gc^\prime| }
\end{array}
\right].
\label{S1}
\end{align}

We select the first $\Csf(\Nsf-t)=\Csf(\Nsf-|\Gc^\prime|)$ rows of $\mathbf{F}$ in form as 
$$\mathbf{F}([\Csf(\Nsf-|\Gc^\prime|)],.):=\begin{bmatrix}
     \mathbf{A}^{\prime}(\{\overline{\Gc^\prime}(1)\},.) \\
       \mathbf{A}^{\prime}(\{\overline{\Gc^\prime}(2)\},.) \\
       \vdots  \\
\mathbf{A}^{\prime}(\{\overline{\Gc^\prime}(\Nsf-|\Gc^\prime|)\},.)
\end{bmatrix},$$
where each submatrix $\mathbf{A}^{\prime}(\{\overline{\Gc^\prime}(i)\},.)$ for each $i\in[\Nsf-|\Gc^\prime|]$ is given by $$\mathbf{A}^{\prime}(\{\overline{\Gc^\prime}(i)\},.)=\left.\begin{bmatrix}
    \mathbf{A}(\{\overline{\Gc^\prime}(i)\},.)\\
    \mathbf{A}(\{\overline{\Gc^\prime}(i)\},.)\\
    \vdots\\
    \mathbf{A}(\{\overline{\Gc^\prime}(i)\},.)
\end{bmatrix} \right\} \quad \text{$\Csf$ copies}.$$

Note that `*' means that the element at this position is selected from the field $\mathbb{F}_{\asf}$, $\asf$ is large enough. 

After this step, the design of the last $\Csf|\Gc^\prime|$ rows of $\mathbf{F}$ aims to ensure the encodability for each worker $n \in \Gc^\prime$. Specifically, for each column $k \in [\Ksf]$, among the last $\Csf|\Gc^\prime|$ elements, we randomly fix $\Csf(|\Gc^\prime|-|\overline{\Cc_{k}}\cap\Gc^\prime|)$ positions and assign values to these positions, then solve the remaining $\Csf|\overline{\Cc_{k}}\cap\Gc^\prime|$ variables by \eqref{computation constraint}. To formalize this, consider the matrix
\begin{align}
    \begin{bmatrix}
\mathbf{A}(\{\Gc^\prime(1)\},.)\\
\mathbf{A}(\{\Gc^\prime(2)\},.)\\
\vdots\\
\mathbf{A}(\{\Gc^\prime(|\Gc^\prime|)\},.)
\end{bmatrix}.
\end{align}
\begin{itemize}
    \item For each row $i\in[|\Gc^\prime|]$, if the element at $\mathbf{A}(\{\Gc^\prime(i)\},\{k\})$ is marked as `*', then the elements in $\mathbf{F}([\Csf(\Nsf-|\Gc^\prime|+i-1)+1:\Csf(\Nsf-|\Gc^\prime|+i)],\{k\})$ correspond to the fixed elements, i.e., $[\Csf(\Nsf-|\Gc^\prime|+i-1)+1:\Csf(\Nsf-|\Gc^\prime|+i)]\subseteq\Fc_k$, and we can select these elements uniformly i.i.d. from the field $\mathbb{F}_{\asf}$. Otherwise, if the element at  $\mathbf{A}(\{\Gc^\prime(i)\},\{k\})$ is marked as `0', then the elements in $\mathbf{F}([\Csf(\Nsf-|\Gc^\prime|+i-1)+1:\Csf(\Nsf-|\Gc^\prime|+i)],\{k\})$ correspond to the variables, i.e., $[\Csf(\Nsf-|\Gc^\prime|+i-1)+1:\Csf(\Nsf-|\Gc^\prime|+i)]\subseteq\Vc_k$. 
    \item Note that if the entry $\mathbf{A}(\{\Gc^\prime(i)\},\{k\})$ is marked as `*', we have $\Gc^\prime(i)\notin\overline{\Cc_{k}}\cap\Gc^\prime$. Recall \eqref{SG design}, the columns with indices in  $[\Csf(\Nsf-|\Gc^\prime|+i-1)+1:\Csf(\Nsf-|\Gc^\prime|+i)]$  in $\mathbf{S}_k^\prime$ are entirely composed of zeros, which correspond to the fixed elements. Therefore, the submatrix of $\mathbf{S}_k^\prime$ formed by selecting columns with indices in $\Fc_k$ is an all-zero matrix, and the submatrix formed by selecting columns with indices in $\Vc_k$ is an identity matrix with dimension $\Csf|\overline{\Cc_{k}}\cap\Gc^\prime|\times \Csf |\overline{\Cc_{k}}\cap\Gc^\prime|$.
\end{itemize}
Then for each $k\in[\Ksf]$ we have
\begin{align*}
\mathbf{S}_k^\prime(.,[\Csf(\Nsf-|\Gc^\prime|)])  \mathbf{F}([\Csf(\Nsf-|\Gc^\prime|)] ,\{k\})+\mathbf{S}_k^\prime(.,\Fc_k)  \mathbf{F}(\Fc_k ,\{k\})+\mathbf{S}_k^\prime(.,\Vc_k)  \mathbf{F}(\Vc_k,\{k\})=\boldsymbol{0}_{\Csf|\overline{\Cc_{k}}\cap\Gc^\prime|\times1},
\end{align*}
where the elements in $\mathbf{S}_k^\prime(.,[\Csf(\Nsf-|\Gc^\prime|)])$ and $\mathbf{S}_k^\prime(.,\Fc_k)$ are all `0', and $\mathbf{S}_k^\prime(.,\Vc_k)$ is an identity matrix. Thus, we can solve $\mathbf{F}(\Vc_k,\{k\})=[0,\ldots,0]^\Tsf$. Hence, we can obtain the whole matrix in form as
\begin{align}
\mathbf{F}=
   \begin{bmatrix}
     \mathbf{A}^\prime(\{\overline{\Gc^\prime}(1)\},.)\\
        \vdots\\
\mathbf{A}^\prime(\{\overline{\Gc^\prime}(\Nsf-|\Gc^\prime|)\},.)\\
\mathbf{A}^\prime(\{\Gc^\prime(1)\},.)\\
\vdots\\
\mathbf{A}^\prime(\{\Gc^\prime(|\Gc^\prime|)\},.)
\end{bmatrix},
\label{solve F1}
\end{align}
 where `0' represents that the element at this position is zero and `*' represents that the element at this position is selected from the field $\mathbb{F}_{\asf}$.
% As a result, each column in \( \mathbf{F}(.,\overline{\Sc_{\overline{\Gc^\prime}(i)}})   \) contains at most $\Nsf-1$ zeros, and no two columns have the same form of $\Nsf-1$ zeros; and no three columns have the same form of $\Nsf-2$ zeros, and so on. Considering that the non-zero elements in \( \mathbf{F}(.,\overline{\Sc_{\overline{\Gc^\prime}(i)}})  \) are uniformly i.i.d. over $\mathbb{F}_{\asf}$, thus \( \mathbf{F}(.,\overline{\Sc_{\overline{\Gc^\prime}(i)}})   \) is column-wise full rank with high probability.As a result, each column in \( \mathbf{F}(.,\overline{\Sc_{\overline{\Gc^\prime}(i)}})   \) contains at most $\Nsf-1$ zeros, and no two columns have the same form of $\Nsf-1$ zeros; and no three columns have the same form of $\Nsf-2$ zeros, and so on. Considering that the non-zero elements in \( \mathbf{F}(.,\overline{\Sc_{\overline{\Gc^\prime}(i)}})  \) are uniformly i.i.d. over $\mathbb{F}_{\asf}$, thus \( \mathbf{F}(.,\overline{\Sc_{\overline{\Gc^\prime}(i)}})   \) is column-wise full rank with high probability.

Next, we aim to prove that \( \mathbf{F}(.,\overline{\Sc_{\overline{\Gc^\prime}(i)}})  \) is column-wise full rank with high probability for each $i\in[|\overline{\Gc^\prime}|]$. Notice that any all-zero submatrix of \( \mathbf{A}(.,\overline{\Sc_{\overline{\Gc^\prime}(i)}})\) consisting of \( |\Gc_1| \) rows and \(|\Qc_1|\) columns satisfies \( \Csf|\Gc_1| + |\Qc_1|\leq \Csf\Nsf\)\footnote{We have \( \mathbf{A}(\{\overline{\Gc^\prime}(i)\},\overline{\Sc_{\overline{\Gc^\prime}(i)}})=[0,\ldots,0]\). Suppose there exists an all-zero submatrix indexed by $(\Gc_1,\Qc_1)$ such that $\Csf|\Gc_1| + |\Qc_1|> \Csf\Nsf$, this necessarily implies $\overline{\Gc^\prime}(i)\notin\Gc_1$. By augmenting the row index set as $\Gc_1:=\Gc_1\cup\{\overline{\Gc^\prime}(i)\}$, the above inequality remains valid, which leads to a contradiction that $\overline{\Gc^\prime}(i)\in\Gc^\prime$. Therefore, any all-zero submatrix of \( \mathbf{A}(.,\overline{\Sc_{\overline{\Gc^\prime}(i)}})\) consisting of \( |\Gc_1| \) rows and \(|\Qc_1|\) columns must satisfy \( \Csf|\Gc_1| + |\Qc_1|\leq \Csf\Nsf\).}, given the structured construction in \eqref{solve F1}, each row of $\mathbf{A}$ is repeated $\Csf$ times along the rows, thus any all-zero submatrix consisting of \( |\Gc_2| \) rows where $\Gc_2=\{[\Csf(i-1)+1:\Csf i]\ | \ i\in\Gc_1\}$
and \(|\Qc_2|\) columns where $\Qc_2=\Qc_1$ in $\mathbf{F}(.,\overline{\Sc_{\overline{\Gc^\prime}(i)}}) $ satisfies \( |\Gc_2| + |\Qc_2|\leq \Csf\Nsf\).

For each column \(j\in\overline{\Sc_{\overline{\Gc^\prime}(i)}}\), define the set of row indices corresponding to non-zero entries as
\[
\mathcal Z_j \triangleq \{\, r\in[\Csf\Nsf] : \mathbf{F}(\{r\},\{j\})\neq 0 \,\}.
\]
By the structural constraint that any all-zero submatrix of \( \mathbf{F}(.,\overline{\Sc_{\overline{\Gc^\prime}(i)}})  \) with \(|\Gc_2|\) rows and \(|\Qc_2|\) columns satisfies
\(|\Gc_2|+|\Qc_2|\le \Csf\Nsf\),
it follows that for any $\Bc\subseteq\overline{\Sc_{\overline{\Gc^\prime}(i)}}$ and $|\Bc|=k$, we have
\[
\left|\bigcap_{j\in\Bc} \mathcal Z_j^c\right|\le \Csf\Nsf-k,
\]
or equivalently,
\[
\left|\bigcup_{j\in\Bc} \mathcal Z_j\right|\ge k.
\]
Based on Hall's Marriage Theorem \cite{hall1987representatives}, we can always select \(|\overline{\Sc_{\overline{\Gc^\prime}(i)}}|\) pairwise distinct row indices
\(z_1,\dots,z_{|\overline{\Sc_{\overline{\Gc^\prime}(i)}}|}\) such that \(z_j\in\mathcal Z_j\) for all \(j\in[|\overline{\Sc_{\overline{\Gc^\prime}(i)}}|]\).
Thus, there exists an \(|\overline{\Sc_{\overline{\Gc^\prime}(i)}}|\times |\overline{\Sc_{\overline{\Gc^\prime}(i)}}|\) submatrix whose diagonal entries are all `*', where `*' represents that the element at this position is selected from the field $\mathbb{F}_{\asf}$.
Hence, the determinant of this submatrix is a non-zero multivariate polynomial   with high probability,
and thus
\(\mathbf F(.,\overline{\Sc_{\overline{\Gc^\prime}(i)}})\)
is column-wise full rank with high probability.

 Note that for each $i\in[|\overline{\Gc^\prime}|]$, we have \( \mathbf{F}([\Csf(i-1)+1:\Csf i],\overline{\Sc_{\overline{\Gc^\prime}(i)}}) =\begin{bmatrix}
0&\ldots&0 \\
    \vdots  & \ddots & \vdots\\
    0&\ldots&0
\end{bmatrix}\), we let $\mathbf{S}_{\overline{\Gc^\prime}(i)}(.,[\Csf(i-1)+1:\Csf i])=\mathbf{I}_{\Csf\times\Csf}$, a full-rank sub-matrix of dimension $|\overline{\Sc_{\overline{\Gc^\prime}(i)}}|\times|\overline{\Sc_{\overline{\Gc^\prime}(i)}}|$ can be obtained by selecting rows with indices in $\Dc=\{z_1,\dots,z_{|\overline{\Sc_{\overline{\Gc^\prime}(i)}}|}\}$, where $\Dc\subseteq[\Csf\Nsf] \setminus [\Csf(i-1)+1:\Csf i]$.
Thus, we can further let \( \mathbf{S}_{\overline{\Gc^\prime}(i)}(.,[\Csf\Nsf] \setminus ([\Csf(i-1)+1:\Csf i]\cup\Dc)) = \boldsymbol{0} \), leading to    
 \begin{align}
\mathbf{S}_{\overline{\Gc^\prime}(i)}(.,\Dc) \mathbf{F}(\Dc,\overline{\Sc_{\overline{\Gc^\prime}(i)}})= -\mathbf{F}([\Csf(i-1)+1:\Csf i],\overline{\Sc_{\overline{\Gc^\prime}(i)}})=\boldsymbol{0},
\end{align}
then we have $\mathbf{S}_{\overline{\Gc^\prime}(i)}= \begin{bmatrix}
0 & \cdots & 0  & 1 & 0 & \cdots &0  & 0 & \cdots & 0\\
0  & \cdots & 0 & 0 & 1 & \cdots & 0 & 0 & \cdots & 0\\
\vdots  & \ddots & \vdots & \vdots & \vdots & \ddots & \vdots & \vdots  & \ddots & \vdots\\
0  & \cdots & 0 &  0 & 0 & \cdots & 1  & 0 & \cdots & 0
    \end{bmatrix}$, where $\mathbf{S}_{\overline{\Gc^\prime}(i)}(.,[\Csf(i-1)+1:\Csf i])=\mathbf{I}_{\Csf\times \Csf}.$

By the construction, we have
\begin{align}
\begin{bmatrix}
    \mathbf{S}_{\overline{\Gc^\prime}(1)}\\
    \vdots\\
    \mathbf{S}_{\overline{\Gc^\prime}(\Nsf-|\Gc^\prime|)}\\
     \mathbf{S}_{\Gc^\prime(1)}\\
    \vdots\\
\mathbf{S}_{\Gc^\prime(|\Gc^\prime|)}
\end{bmatrix}
=\begin{bmatrix}
1 & 0 & 0 & \cdots & 0 \\
0 & 1 & 0 & \cdots & 0 \\
0 & 0 & 1 & \cdots & 0 \\
\vdots & \vdots & \vdots & \ddots & \vdots \\
0 & 0 & 0 & \cdots & 1
\end{bmatrix},
\end{align}
which is an identity matrix and is thus full-rank.

(ii) $|\Gc^\prime|>t$. For any set $\Vc_k \subseteq [\Csf(\Nsf - t)+1 : \Csf \Nsf]$ with $|\Vc_k| = \Csf |\overline{\Cc_k} \cap \Gc^\prime|$, the columns are selected from the last $\Csf t$ columns.
When $|\Gc^\prime| > t$, the submatrix $\mathbf{S}_k^\prime(., \Vc_k)$ is not necessarily full rank  under the construction in \eqref{SG design},
since the columns indexed by $[\Csf(\Nsf-|\Gc^\prime|)+1:\Csf(\Nsf-t)]$ consist entirely of zero entries.

We select
    \begin{align}
     \begin{bmatrix}
           \mathbf{S}_{\Gc^\prime(1)}\\
            \mathbf{S}_{\Gc^\prime(2)}\\
            \ldots\\
    \mathbf{S}_{\Gc^\prime(|\Gc^\prime|)}
        \end{bmatrix}=
    \left[
    \begin{array}{c@{\hskip 6pt}c@{\hskip 6pt}c}
    \boldsymbol{0}_{\Csf|\Gc^\prime|\times\Csf(\Nsf-|\Gc^\prime|)} & \vdashline&  \mathbf{M}_{\Csf|\Gc^\prime|\times\Csf|\Gc^\prime| }
    \end{array}
    \right],
    \label{S2}
    \end{align}
    where $\mathbf{M}$ is an MDS matrix. Thus, for any set $\Vc_k\subseteq[\Csf(\Nsf-t)+1:\Csf\Nsf]$, $|\Vc_k|=\Csf|\overline{\Cc_{k}}\cap\Gc^\prime|$, the submatrix $\mathbf{S}_k^\prime(.,\Vc_k)$ is full rank. Therefore, next we only need to construct a matrix $\mathbf{F}$ that satisfies the encodability of each worker $n\in\Gc^\prime$ under the selection of $ \begin{bmatrix}
        \mathbf{S}_{\Gc^\prime(1)}\\
       \mathbf{S}_{\Gc^\prime(2)}\\
        \ldots\\
\mathbf{S}_{\Gc^\prime(|\Gc^\prime|)}
    \end{bmatrix}$ in \eqref{S2}.
As demonstrated for the selection in \eqref{S1}, the matrix $\mathbf{F}$ constructed in \eqref{solve F1} satisfies the encodability constraint. Building on this result, a similar construction can be applied under the selection in \eqref{S2}, yielding:
\begin{align}
\mathbf{F}=&\begin{bmatrix}
\mathbf{I}_{\Csf(\Nsf-|\Gc^\prime|)\times\Csf(\Nsf-|\Gc^\prime|)} & \boldsymbol{0}_{\Csf(\Nsf-|\Gc^\prime|)\times\Csf|\Gc^\prime|}\\
\boldsymbol{0}_{\Csf|\Gc^\prime|\times\Csf(\Nsf-|\Gc^\prime|)} & \mathbf{M}^{-1}_{\Csf|\Gc^\prime|\times\Csf|\Gc^\prime|
}
\end{bmatrix}  \begin{bmatrix}
     \mathbf{A}^\prime(\{\overline{\Gc^\prime}(1)\},.)\\
        \vdots\\
\mathbf{A}^\prime(\{\overline{\Gc^\prime}(\Nsf-|\Gc^\prime|)\},.)\\
\mathbf{A}^\prime(\{\Gc^\prime(1)\},.)\\
\vdots\\
\mathbf{A}^\prime(\{\Gc^\prime(|\Gc^\prime|)\},.)
\end{bmatrix},
% \nonumber \\
% =&\begin{bmatrix}
% \mathbf{I}_{(\Nsf-|\Gc^\prime|)\times(\Nsf-|\Gc^\prime|)} & \boldsymbol{0}_{(\Nsf-|\Gc^\prime|)\times|\Gc^\prime|}\\
% \boldsymbol{0}_{|\Gc^\prime|\times(\Nsf-|\Gc^\prime|)} & \mathbf{M}^{-1}_{|\Gc^\prime|\times|\Gc^\prime|}
% \end{bmatrix}\begin{bmatrix}
%     \mathbf{B}_{(\Nsf-|\Gc^\prime|)\times\Ksf}\\
% \mathbf{C}_{|\Gc^\prime|\times\Ksf}
% \end{bmatrix}\nonumber \\
% =&\begin{bmatrix}
%     \mathbf{B}_{(\Nsf-|\Gc^\prime|)\times\Ksf}\\
% \mathbf{M}^{-1}_{|\Gc^\prime|\times|\Gc^\prime|}\mathbf{C}_{|\Gc^\prime|\times\Ksf}
% \end{bmatrix}.
\end{align}
the encodability constraint is satisfied because
\begin{align}
 \begin{bmatrix}
        \mathbf{S}_{\Gc^\prime(1)}\\
       \mathbf{S}_{\Gc^\prime(2)}\\
        \ldots\\
\mathbf{S}_{\Gc^\prime(|\Gc^\prime|)}
    \end{bmatrix}\mathbf{F}=\begin{bmatrix}
\mathbf{A}^\prime(\{\Gc^\prime(1)\},.)\\
\vdots\\
\mathbf{A}^\prime(\{\Gc^\prime(|\Gc^\prime|)\},.)
\end{bmatrix}.
\end{align}

We have shown that for each $i\in[|\overline{\Gc^\prime}|]$, the columns of matrix $ \begin{bmatrix}
     \mathbf{A}^\prime(\{\overline{\Gc^\prime}(1)\},.)\\
        \vdots\\
\mathbf{A}^\prime(\{\overline{\Gc^\prime}(\Nsf-|\Gc^\prime|)\},.)\\
\mathbf{A}^\prime(\{\Gc^\prime(1)\},.)\\
\vdots\\
\mathbf{A}^\prime(\{\Gc^\prime(|\Gc^\prime|)\},.)
\end{bmatrix}$ indexed by $\overline{\Sc_{\overline{\Gc^\prime}(i)}}$ are linearly independent with high probability.
Thus, the submatrix \( \mathbf{F}(.,\overline{\Sc_{\overline{\Gc^\prime}(i)}})  \) is column-wise full rank with high probability, since the matrix $\begin{bmatrix}
\mathbf{I}_{\Csf(\Nsf-|\Gc^\prime|)\times\Csf(\Nsf-|\Gc^\prime|)} & \boldsymbol{0}_{\Csf(\Nsf-|\Gc^\prime|)\times\Csf|\Gc^\prime|}\\
\boldsymbol{0}_{\Csf|\Gc^\prime|\times\Csf(\Nsf-|\Gc^\prime|)} & \mathbf{M}^{-1}_{\Csf|\Gc^\prime|\times\Csf|\Gc^\prime|
}
\end{bmatrix}$ is full rank.

For each $i\in[|\overline{\Gc^\prime}|]$, \( \mathbf{F}( [\Csf (i-1)+1:\Csf i],\overline{\Sc_{\overline{\Gc^\prime}(i)}}) =\boldsymbol{0} \), we can let $\mathbf{S}_{\overline{\Gc^\prime}(i)}(.,[\Csf(i-1)+1:\Csf i])=\mathbf{I}_{\Csf\times\Csf}$ and a full-rank sub-matrix of \( \mathbf{F}(.,\overline{\Sc_{\overline{\Gc^\prime}(i)}})  \) with dimension $|\overline{\Sc_{\overline{\Gc^\prime}(i)}}|\times|\overline{\Sc_{\overline{\Gc^\prime}(i)}}|$ can be obtained by selecting rows with indices in $\Dc$, where $|\Dc|=|\overline{\Sc_{\overline{\Gc^\prime}(i)}}|$ and $\Dc\subseteq[\Csf\Nsf] \setminus [\Csf(i-1)+1:\Csf i]$.
We then set \( \mathbf{S}_{\overline{\Gc^\prime}(i)}(.,[\Csf\Nsf] \setminus ([\Csf(i-1)+1:\Csf i]\cup\Dc)) = \boldsymbol{0} \) to have  
 \begin{align}
\mathbf{S}_{\overline{\Gc^\prime}(i)}(.,\Dc) \mathbf{F}(\Dc,\overline{\Sc_{\overline{\Gc^\prime}(i)}})= -\mathbf{F}([\Csf(i-1)+1:\Csf i],\overline{\Sc_{\overline{\Gc^\prime}(i)}})=\boldsymbol{0},
\end{align}
we have $\mathbf{S}_{\overline{\Gc^\prime}(i)}= \begin{bmatrix}
0 & \cdots & 0  & 1 & 0 & \cdots &0  & 0 & \cdots & 0\\
0  & \cdots & 0 & 0 & 1 & \cdots & 0 & 0 & \cdots & 0\\
\vdots  & \ddots & \vdots & \vdots & \vdots & \ddots & \vdots & \vdots  & \ddots & \vdots\\
0  & \cdots & 0 &  0 & 0 & \cdots & 1  & 0 & \cdots & 0
    \end{bmatrix}$, where $\mathbf{S}_{\overline{\Gc^\prime}(i)}(.,[\Csf(i-1)+1:\Csf i])=\mathbf{I}_{\Csf\times \Csf}.$

By the construction, we have
\begin{align}
\begin{bmatrix}
    \mathbf{S}_{\overline{\Gc^\prime}(1)}\\
    \vdots\\
    \mathbf{S}_{\overline{\Gc^\prime}(\Nsf-|\Gc^\prime|)}\\
    \mathbf{S}_{\Gc^\prime(1)}\\
    \vdots\\
\mathbf{S}_{\Gc^\prime(|\Gc^\prime|)}
\end{bmatrix}
=\begin{bmatrix}
\mathbf{I}_{\Csf(\Nsf-|\Gc^\prime|)\times\Csf(\Nsf-|\Gc^\prime|)} & \boldsymbol{0}_{\Csf(\Nsf-|\Gc^\prime|)\times\Csf|\Gc^\prime|}\\
\boldsymbol{0}_{\Csf|\Gc^\prime|\times\Csf(\Nsf-|\Gc^\prime|)} & \mathbf{M}_{\Csf|\Gc^\prime|\times\Csf|\Gc^\prime|}
\end{bmatrix},
\end{align}
which is full-rank.

As a result, we proved that $P_{n}^\prime$ and $T^\prime$ are non-zero multivariate polynomials.  From \eqref{solution for the Cramer's rule2}, it can be seen that each element in \( \{ \mathbf{S}_n:n\in\overline{\Gc^\prime}\} \) has degree no more than \( 2\Csf|\overline{\Cc_{k}}\cap\Gc^\prime||\overline{\Sc_{n}}|+|\overline{\Sc_{n}}| \), whose variables are the elements in $\{ \mathbf{S}_{n}:n\in\Gc^\prime\}$ and the randomly selected elements in ${\mathbf{F}}(\Fc_k\cup[\Csf(\Nsf-t)],\{k\})$ for each $k\in[\Ksf]$ and $ \mathbf{S}_n(.,[\Nsf]\setminus\Dc)$ for each $n\in\overline{\Gc^\prime}$. In addition, each term in \(\mathbf{T} \) is a multivariate polynomial whose variables are the elements in \(\mathbf{T}\) with degree \( \Csf\Nsf \). Hence, \( T_i \) and \( K_i \) are multivariate polynomials  with degree no more than \(  (2\Csf|\overline{\Cc_{k}}\cap\Gc^\prime|+1)|\overline{\Sc_{n}}|\Csf\Nsf \),  whose variables are the elements in $\{\mathbf{S}_{n}:n\in\Gc^\prime\}$ and the randomly selected elements in ${\mathbf{F}}(\Fc_k\cup[\Csf(\Nsf-t)],\{k\})$ for each $k\in[\Ksf]$ and $\mathbf{S}_n(.,[\Nsf]\setminus\Dc)$ for each $n\in\overline{\Gc^\prime}$.
Then
 by the Schwartz--Zippel Lemma~\cite{Schwartz,Zippel,Demillo_Lipton}, we have
\begin{align}
\Pr \{T^\prime \neq 0\} \geq 1 - \frac{(\Csf\Nsf)! (2\Csf|\overline{\Cc_{k}}\cap\Gc^\prime|+1)|\overline{\Sc_{n}}|\Csf\Nsf}{\asf}. \label{Pknoteq02}
\end{align}

Thus, we have
\begin{subequations}
    \begin{align}
&\Pr \{\mathbf{T} \text{ is full-rank}\} 
\geq 1 - \Pr \{\mathbf{T} \text{ does not exist}\} - \Pr \{T^\prime = 0\}  \\
&\geq 1 - |\overline{\Gc^\prime}|\left( \frac{|\overline{\Sc_{n}}|\Csf^2|\overline{\Cc_{k}}\cap\Gc^\prime|^2}{\asf}+\frac{2\binom{\Csf\Nsf}{|\overline{\Sc_{n}}|}\left(|\overline{\Sc_{n}}|\right)! \Csf|\overline{\Cc_{k}}\cap\Gc^\prime||\overline{\Sc_{n}}|}{\asf}\right)-\frac{(\Csf\Nsf)! (2\Csf|\overline{\Cc_{k}}\cap\Gc^\prime|+1)|\overline{\Sc_{n}}|\Csf\Nsf}{\asf}\\
&\xrightarrow{\asf \to \infty} 1.
\end{align}
\end{subequations}
Therefore, we proved that the matrix in \eqref{the matrix of sk}
has rank equal to $\Csf\Nsf$  with high probability.

\section{Converse Bound for Theorem~\ref{thm:first result}}
\label{sec:proof of converse}
%When we consider the communication cost $\Csf=\frac{p}{q}\Lsf$, for one pair $(\Gc,\Qc) \in \Zc$,
%the converse bound follows directly from the fact that the left null space of the matrix $\mathbf{F}(.,\Qc^{q})$, which has dimension \( p\Nsf_\rsf \times q|\Qc| \), contains at least \(p |\Gc| \) linearly independent vectors (since \(  |\Gc|<\Nsf_\rsf \)). Moreover, the rank of $\mathbf{F}(., \Qc^{q})$ must satisfy
%$\text{rank}(\mathbf{F}(.,\Qc^{q})) \geq \text{min}\{q\Ksf_\csf, q|\Qc|\}$, which leads to $q\Ksf_\csf\leq p\Nsf_\rsf-p |\Gc|$, we can obtain $\Ksf_\csf\leq\lfloor\frac{p}{q}(\Nsf_\rsf- \ \alpha)\rfloor$.

Given the data assignment matrix $\mathbf{A}$, the communication cost for each worker is $\Csf$, we define the set:
\begin{align}
	\Zc=\{(\Gc,\Qc)&|\Gc\subseteq[\Nsf],\Qc\subseteq[\Ksf],\nonumber \\
	&\mathbf{A}(\Gc,\Qc)=\boldsymbol{0}_{|\Gc|\times|\Qc|},\Csf|\Gc|+|\Qc|>\Csf\Nsf \},
    \label{important set}
\end{align} where each pair $(\Gc,\Qc) \in \Zc$  indicates that no dataset in  $\Qc$ is stored by any worker in $\Gc$. 
Since the elements of $\mathbf{F}_1$ are uniformly i.i.d. over the field $\mathbb{F}_{\asf}$, to recover the corresponding coded information of datasets in $\Qc$ with $\min \{\Ksf_\csf,|\Qc|\}\Lsf$ symbols, the total amount of information transmitted from the remaining $\Nsf - |\Gc|$ workers must be at least $\min \{\Ksf_\csf,|\Qc|\}\Lsf$, thus 
\begin{align}
	(\Nsf-|\Gc|)\Csf\Lsf\geq\min \{\Ksf_\csf,|\Qc|\}\Lsf.
\end{align}
Given that $\Csf|\Gc|+|\Qc|>\Csf\Nsf$,
this implies $\Ksf_\csf\leq (\Nsf-|\Gc|)\Csf$. Recall that we define  $\alpha = \max \left\{ \left| \Gc \right| \,\middle|\, (\Gc, \Qc) \in \Zc \right\}$ and $\Ksf_\csf\leq\Ksf$,
we can obtain $$\Ksf_\csf\leq\min\{\Csf(\Nsf- \ \alpha),\Ksf\}.$$

\section{
The General Scheme for Fractional Communication Costs}
\label{subsec2:the scheme given an integer communication cost}
Under the no-subpacketization assumption, we previously developed a scheme simultaneously satisfying encodability and decodability constraints. In this section, we generalize the scheme to fractional communication costs $\Csf=\frac{p}{q}$ for any $p,q\in\mathbb{N}^+$ with $\gcd(p,q)=1$, where $0\leq \Csf \leq \max_{n\in[\Nsf]} |\Sc_n|$. When $q=1$, the generalized scheme reduces to the one presented in Section~\ref{subsec:the scheme under unit communication cost}. We now proceed to describe the construction of the generalized scheme for $q>1$.
 
For each $W_k$ where $k\in[\Ksf]$, we divide it into $q$ non-overlapping and equal-length pieces, denoted as $W_k=(W_{k,1},W_{k,2},\ldots,W_{k,q})$. The computational task can be expressed as 
\begin{align}
\mathbf{F}_1\begin{bmatrix}
            W_{1,1}\\ 
             W_{2,1}\\
             \vdots\\
              W_{\Ksf,1}\\
               W_{1,2}\\
               \vdots\\
                W_{\Ksf,q}
        \end{bmatrix},
        \label{task3}
\end{align} 
which contains $q\Ksf_\csf$ linearly independent combinations of the pieces $\{ W_{1,1},\ldots, W_{\Ksf,q}\}$, corresponding to $\Ksf_\csf \Lsf$ symbols of information. After receiving the messages from the workers, the server can recover $p\Nsf$-dimensional linear combinations of these pieces, where $\mathbf{F}_2$ represents the virtual
demand coefficient matrix with dimension $(p\Nsf-q\Ksf_\csf)\times q\Ksf$.

We now proceed to provide theoretical bounds on the converse and achievable bounds.

First, we define the sparsity property of the matrix $\mathbf{A}$ under a fixed communication cost $\Csf=\frac{p}{q}$, which will be used in the proposed converse bound and achievable scheme. Let 
    \begin{align}
    	\Zc=\{(\Gc,\Qc)&|\Gc\subseteq[\Nsf],\Qc\subseteq[\Ksf],\nonumber \\
    	&\mathbf{A}(\Gc,\Qc)=\boldsymbol{0}_{|\Gc|\times|\Qc|},\Csf|\Gc|+|\Qc|>\Csf\Nsf \}.
    	\label{the sparsity of A for fractional communication costs}
    \end{align}
    
This follows from that in  the assumption that the submatrix $\mathbf{A}(\Gc, \Qc)$ is all-zero, which implies that none of the workers in $\Gc$ store any datasets from $\Qc$. Consequently, for the pieces $W_k=(W_{k,1},\ldots, W_{k,q})$ with $k\in\Qc$,  the total dimension of information that the master can receive from the remaining $\Nsf-|\Gc|$ workers is at most $p(\Nsf-|\Gc|)$. Since $p(\Nsf-|\Gc|)<q|\Qc|$, which follows from the condition $\Csf|\Gc|+|\Qc|>\Csf\Nsf$, this amount of information is insufficient to recover the data corresponding to the datasets in $\Qc$. Thus, the set $\Zc$ identifies the all-zero submatrices in the data assignment matrix, each of which limits communication efficiency in distributed computing.

\begin{proposition}[Converse bound] 
\label{proposition:first result}
For the $(\Ksf,\Nsf,\Csf)$ heterogeneous distributed system with arbitrary data assignment and a fixed communication cost  $0\leq\Csf \leq \max_{n\in[\Nsf]} |\Sc_n|$, let $\alpha = \max \left\{ \left| \Gc \right| \,\middle|\, (\Gc, \Qc) \in \Zc \right\}$, the maximum computable dimension should satisfy
\begin{align}
\Ksf_\csf\leq \min \{\Csf(\Nsf- \ \alpha),\Ksf\} .  \label{proposition:now first result}
\end{align}
\end{proposition}

\begin{proposition}[Achievable bound] 
\label{proposition:second result}
For the $(\Ksf,\Nsf,\Csf)$ heterogeneous distributed system with arbitrary data assignment and a fixed communication cost $0\leq\Csf \leq \max_{n\in[\Nsf]} |\Sc_n|$, let $\Gc^\prime=\bigcup\limits_{(\Gc,\Qc)\in\Zc} \Gc$ and \(t \) represent the maximum number of `0' in each column of the sub-matrix $\mathbf{A}(\Gc^\prime,.)$, the computable dimension is achievable
\begin{align}
\Ksf_\csf=\min \{\Csf(\Nsf-t),\Ksf \}.  \label{proposition:now second result}
\end{align}
\end{proposition}

We let each worker send
\begin{align}
        X_n=\mathbf{S}_n \mathbf{F}\begin{bmatrix}
            W_{1,1}\\ 
             \vdots\\
              W_{\Ksf,1}\\
               W_{1,2}\\
               \vdots\\
                W_{\Ksf,q}
        \end{bmatrix}=\mathbf{S}_n \begin{bmatrix}
           \mathbf{F}_1\\ 
              \mathbf{F}_2\\
        \end{bmatrix}\begin{bmatrix}
            W_{1,1}\\ 
             \vdots\\
              W_{\Ksf,1}\\
               W_{1,2}\\
               \vdots\\
                W_{\Ksf,q}
        \end{bmatrix},
        \label{the transmission of worker n2}
    \end{align}
each worker sends $p$ linear combinations  of pieces, the transmitted message $X_n$ contains $\frac{p}{q}\Lsf$ symbols.  We define $\overline{\Sc_n}^{q}=\bigcup_{i\in[0:q-1]}(\overline{\Sc_n}+i\Ksf)$ for each $n\in[\Nsf]$. By the same way, we have the following two constraints:
\begin{constraint}[Encodability constraint] 
    \label{pro:encoding2} 
   To ensure that each worker only participates in computations related to the datasets it owns, for each worker $n\in[\Nsf]$ we have 
    \begin{align}
\mathbf{S}_n\mathbf{F}(.,\overline{\Sc_{n}}^{q})=\boldsymbol{0}_{p\times|\overline{\Sc_{n}}^{q}|}.
     \label{encodability2}
     \end{align}
    \end{constraint}
    \begin{constraint}[Decodability constraint] 
    \label{pro:decoding2} 
    To guarantee that the master can recover the desired computation result,  the matrix
 $ \begin{bmatrix}
       \mathbf{S}_{1} \\
       \vdots \\
       \mathbf{S}_{\Nsf}
    \end{bmatrix}$
     with dimension $p\Nsf\times p\Nsf$ is full rank.
    \end{constraint}
 Thus, the master recovers $\mathbf{F}\mathbf{W}$ by computing 
    \begin{align}
    \label{the recovery of F2}
        \mathbf{F}\mathbf{W}=\begin{bmatrix}
       \mathbf{S}_{1} \\
       \vdots \\
       \mathbf{S}_{\Nsf}
    \end{bmatrix} ^{-1} \begin{bmatrix}
       X_{1} \\
       \vdots \\
       X_{\Nsf}
    \end{bmatrix}.
    \end{align}
From the first $q\Ksf_\csf$ rows of the recovered result, the master can recover the computational task. 

Following the main idea from Section~\ref{subsec:the scheme under unit communication cost}, we now present the selection that satisfies Constraints~\ref{pro:encoding2}-\ref{pro:decoding2}.
The general scheme for the selection of $(\mathbf{S}_n:n\in[\Nsf])$ and $\mathbf{F}_2$ is given as follows:
\begin{itemize}
     \item First, for the data assignment matrix \(\mathbf{A}\), we define 
    \begin{align}
    	\Zc=\{(\Gc,\Qc)&|\Gc\subseteq[\Nsf],\Qc\subseteq[\Ksf],\nonumber \\
    	&\mathbf{A}(\Gc,\Qc)=\boldsymbol{0}_{|\Gc|\times|\Qc|}, \Csf|\Gc|+|\Qc|>\Csf\Nsf \},\nonumber
    \end{align}
    and recall $\alpha = \max \left\{ \left| \Gc \right| \,\middle|\, (\Gc, \Qc) \in \Zc \right\}$, we have the converse $\Ksf_\csf\leq \min \{\Csf(\Nsf-\alpha),\Ksf\}$.   
    \item Then we consider the sub-matrix $\mathbf{A}(\Gc^\prime,.)$ where $\Gc^\prime=\bigcup\limits_{(\Gc,\Qc)\in\Zc} \Gc$. Recall that \( t \) represents the maximum number of `0' in each column of the sub-matrix $\mathbf{A}(\Gc^\prime,.)$. We generate a $p|\Gc^\prime|\times p\Nsf$ matrix with elements uniformly i.i.d. over $\mathbb{F}_{\asf}$ as $[\mathbf{S}_{\Gc^\prime(1)};\mathbf{S}_{\Gc^\prime(2)};\ldots;\mathbf{S}_{\Gc^\prime(|\Gc^\prime|)}]$. For each $l\in[q],k\in[\Ksf]$, the first $q\Ksf_\csf$ elements in ${\mathbf{F}}(.,\{(l-1)\Ksf+k\})$ are predetermined, and the remaining elements should satisfy  
    \begin{align}
        \mathbf{S}_n{\mathbf{F}}(.,\{(l-1)\Ksf+k\})=\boldsymbol{0}_{p\times1}, \forall n\in \overline{\Cc_{k}}\cap\Gc^\prime.
        \label{computation constraint2}
    \end{align}
     The number of such workers, $|\overline{\Cc_{k}}\cap\Gc^\prime|$, corresponds to the number of `0' in  $\mathbf{A}(\Gc^\prime,\{k\})$. We randomly select $p (t-|\overline{\Cc_{k}}\cap\Gc^\prime|)$ positions from the last $p t$ rows and assign values to these positions with elements uniformly i.i.d. over $\mathbb{F}_{\asf}$. Then,  the remaining $p|\overline{\Cc_{k}}\cap\Gc^\prime|$ elements in the column ${\mathbf{F}}(.,\{(l-1)\Ksf+k\})$ are determined by solving the linear constraints specified in~\eqref{computation constraint2}. Thus, we construct the whole matrix $\mathbf{F}$.
    \item Finally, for each worker $n\in[\Nsf]\setminus\Gc^\prime$, we form $\mathbf{S}_n$ by selecting $p$ linearly independent vectors from the left null space of ${\mathbf{F}}(.,\overline{\Sc_{n}}^{q})$. This selection is feasible because for $n\in[\Nsf]\setminus\Gc^\prime$, row $n$ of $\mathbf{A}$ satisfies $\Csf|\Gc|+|\Qc|\leq \Csf\Nsf$ when $|\Gc|=1$, we have 
 \begin{align}
q|\Qc|=q|\overline{\Sc_{n}}|\leq p(\Nsf-1),
\label{the sparsity2}
    \end{align}
   based on this inequality, we have
    \begin{align}
   p\Nsf-|\overline{\Sc_{n}}^{q}|=p\Nsf-q |\overline{\Sc_{n}}|
     \geq p\Nsf-p(\Nsf-1)=p.
     \label{existence2}
       \end{align}
\end{itemize}
Following the construction of the matrix 
$\mathbf{F}$, the last $p t$ rows are partially determined by solving the linear equations imposed by the constraint in~\eqref{computation constraint2}. As a result, only the first $p(\Nsf-t)$ rows of 
$\mathbf{F}$ can be freely chosen, with elements uniformly i.i.d. over $\mathbb{F}_{\asf}$. Hence, the rank of $\mathbf{F}_1$ is at most $p(\Nsf-t)$. Since $\mathbf{F}_1$ has dimension $q\Ksf_\csf \times q\Ksf$, we obtain $q\Ksf_\csf \leq p(\Nsf-t)$, i.e., the proposed scheme achieves the computable dimension $\Ksf_\csf=\min\{ \Csf(\Nsf-t),\Ksf\}$.
\begin{figure} 
  \centering
    \centering
    \includegraphics[scale=0.5]{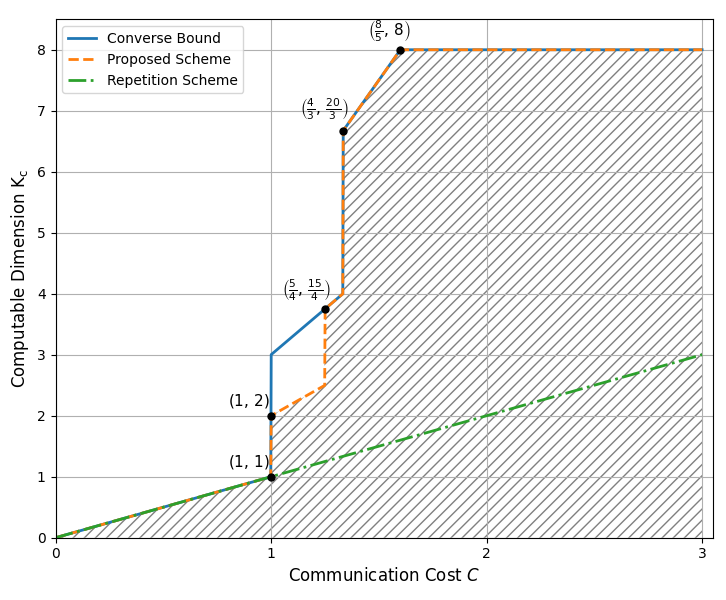}
  \caption{Comparison with the data assignment matrix in Example~\ref{ex851} for fractional communication costs.}
  \label{fig:myfig2}
\end{figure}

We can equivalently view the system as having an integer communication cost $\Csf = p$ and an extended storage pattern $\overline{\Sc_n}^{q} = \bigcup_{i\in[0:q-1]}(\overline{\Sc_n}+i\Ksf)$ for each $n\in[\Nsf]$. Under this perspective, the scheme becomes structurally identical to the integer $\Csf$ case presented in Section~\ref{subsec:the scheme under unit communication cost}. Constraints~\ref{pro:encoding2}-\ref{pro:decoding2} are then direct reformulations of Constraints~\ref{pro:encoding}-\ref{pro:decoding}, with the substitutions $\Csf \rightarrow p$ and $\overline{\Sc_n} \rightarrow \overline{\Sc_n}^{q}$. Consequently, the proof of Constraints~\ref{pro:encoding2}-\ref{pro:decoding2} follows immediately from the proof of Constraints~\ref{pro:encoding}-\ref{pro:decoding}, and we omit the repetitive details for brevity.

Based on the data assignment in Example~\ref{ex851}, we further explore the tradeoff between the communication cost and the computable dimension under fractional communication costs for three cases: the converse bound, the proposed scheme, and the benchmark scheme which repeats the scheme in~\cite{jahani2021optimal} $\Ksf_{\rm c}$ times (referred to as the Repetition Scheme). This is presented in Fig.~\ref{fig:myfig2}. 

The parameters $t$ and $\alpha$ are determined by the combinatorial structure of $\mathbf{A}$ together with the communication constraint $\Csf$. As the communication load $\Csf$ varies, $t$ and $\alpha$ are piecewise constant functions of $\Csf$, exhibiting a stepwise behavior. This piecewise variation leads to the changes in the slope observed in the figure. The shaded region on the right-hand side of the curve for the proposed scheme represents its achievable region. As shown in the figure, the proposed scheme achieves a higher computable dimension than the benchmark scheme under the same communication cost.

\section{conclusions}
\label{sec:conclusion}
This paper studied the distributed linearly separable computation problem with arbitrary heterogeneous data assignment. Focusing on the problem of maximizing the computable dimension subject to a fixed communication cost, we established universal achievable and converse bounds that hold for arbitrary heterogeneous data assignment, for both integer and fractional communication costs. Furthermore, the achievable and converse bounds coincide under some parameter regimes.

Ongoing work includes deriving tighter achievable and converse bounds, extending the framework to incorporate stragglers, and improving communication efficiency under lossy computation.

\appendices 
\bibliographystyle{IEEEtran}
\bibliography{reference}

@article{dean2008mapreduce,
  title={MapReduce: simplified data processing on large clusters},
  author={Dean, Jeffrey and Ghemawat, Sanjay},
  journal={Communications of the ACM},
  volume={51},
  number={1},
  pages={107--113},
  year={2008},
  publisher={ACM New York, NY, USA}
}

@inproceedings{zaharia2010spark,
  title={Spark: Cluster computing with working sets},
  author={Zaharia, Matei and Chowdhury, Mosharaf and Franklin, Michael J and Shenker, Scott and Stoica, Ion},
  booktitle={2nd USENIX workshop on hot topics in cloud computing (HotCloud 10)},
  year={2010}
}

@article{dean2012large,
  title={Large scale distributed deep networks},
  author={Dean, Jeffrey and Corrado, Greg and Monga, Rajat and Chen, Kai and Devin, Matthieu and Mao, Mark and Ranzato, Marc'aurelio and Senior, Andrew and Tucker, Paul and Yang, Ke and others},
  journal={Advances in neural information processing systems},
  volume={25},
  year={2012}
}

@article{jahani2021optimal,
  title={Optimal communication-computation trade-off in heterogeneous gradient coding},
  author={Jahani-Nezhad, Tayyebeh and Maddah-Ali, Mohammad Ali},
  journal={IEEE Journal on Selected Areas in Information Theory},
  volume={2},
  number={3},
  pages={1002--1011},
  year={2021},
  publisher={IEEE}
}

@article{li2014communication,
  title={Communication efficient distributed machine learning with the parameter server},
  author={Li, Mu and Andersen, David G and Smola, Alexander J and Yu, Kai},
  journal={Advances in Neural Information Processing Systems},
  volume={27},
  year={2014}
}

@article{dean2013tail,
  title={The tail at scale},
  author={Dean, Jeffrey and Barroso, Luiz Andr{\'e}},
  journal={Communications of the ACM},
  volume={56},
  number={2},
  pages={74--80},
  year={2013},
  publisher={ACM New York, NY, USA}
}

@inproceedings{li2015coded,
  title={Coded mapreduce},
  author={Li, Songze and Maddah-Ali, Mohammad Ali and Avestimehr, A Salman},
  booktitle={2015 53rd Annual Allerton Conference on Communication, Control, and Computing (Allerton)},
  pages={964--971},
  year={2015},
  organization={IEEE}
}

@inproceedings{huang2023fundamental,
  title={Fundamental limits of distributed linearly separable computation under cyclic assignment},
  author={Huang, Wenbo and Wan, Kai and Sun, Hua and Ji, Mingyue and Qiu, Robert Caiming and Caire, Giuseppe},
  booktitle={2023 IEEE International Symposium on Information Theory (ISIT)},
  pages={2296--2301},
  year={2023},
  organization={IEEE}
}

@inproceedings{gradiencoding,
  title={Gradient coding: Avoiding stragglers in distributed learning},
  author={Tandon, Rashish and Lei, Qi and Dimakis, Alexandros G and Karampatziakis, Nikos},
  booktitle={International Conference on Machine Learning},
  pages={3368--3376},
  year={2017},
  organization={PMLR}
}

@inproceedings{yu2019lagrange,
  title={Lagrange coded computing: Optimal design for resiliency, security, and privacy},
  author={Yu, Qian and Li, Songze and Raviv, Netanel and Kalan, Seyed Mohammadreza Mousavi and Soltanolkotabi, Mahdi and Avestimehr, Salman A},
  booktitle={The 22nd International Conference on Artificial Intelligence and Statistics},
  pages={1215--1225},
  year={2019},
  organization={PMLR}
}

@article{wan2021distributed,
  title={Distributed linearly separable computation},
  author={Wan, Kai and Sun, Hua and Ji, Mingyue and Caire, Giuseppe},
  journal={IEEE Transactions on Information Theory},
  volume={68},
  number={2},
  pages={1259--1278},
  year={2021},
  publisher={IEEE}
}

@inproceedings{halbawi2018improving,
  title={Improving distributed gradient descent using reed-solomon codes},
  author={Halbawi, Wael and Azizan, Navid and Salehi, Fariborz and Hassibi, Babak},
  booktitle={2018 IEEE International Symposium on Information Theory (ISIT)},
  pages={2027--2031},
  year={2018},
  organization={IEEE}
}

@article{Schwartz,
  title={Fast probabilistic algorithms for verification of polynomial identities},
  author={Schwartz, Jacob T},
  journal={Journal of the ACM (JACM)},
  volume={27},
  number={4},
  pages={701--717},
  year={1980},
  publisher={ACM}
}

@inproceedings{Zippel,
  title={Probabilistic algorithms for sparse polynomials},
  author={Zippel, Richard},
  booktitle={International symposium on symbolic and algebraic manipulation},
  pages={216--226},
  year={1979},
  organization={Springer}
}

@article{Demillo_Lipton,
  title={A probabilistic remark on algebraic program testing},
  author={Demillo, Richard A and Lipton, Richard J},
  journal={Information Processing Letters},
  volume={7},
  number={4},
  pages={193--195},
  year={1978},
  publisher={Elsevier}
}

@inproceedings{lee2017high,
	title={High-dimensional coded matrix multiplication},
	author={Lee, Kangwook and Suh, Changho and Ramchandran, Kannan},
	booktitle={2017 IEEE International Symposium on Information Theory (ISIT)},
	pages={2418--2422},
	year={2017},
	organization={IEEE}
}

@article{dutta2019optimal,
	title={On the optimal recovery threshold of coded matrix multiplication},
	author={Dutta, Sanghamitra and Fahim, Mohammad and Haddadpour, Farzin and Jeong, Haewon and Cadambe, Viveck and Grover, Pulkit},
	journal={IEEE Transactions on Information Theory},
	volume={66},
	number={1},
	pages={278--301},
	year={2019},
	publisher={IEEE}
}

@article{dutta2016short,
	title={Short-dot: Computing large linear transforms distributedly using coded short dot products},
	author={Dutta, Sanghamitra and Cadambe, Viveck and Grover, Pulkit},
	journal={Advances In Neural Information Processing Systems},
	volume={29},
	year={2016}
}

@article{cao2021adaptive,
	title={Adaptive gradient coding},
	author={Cao, Hankun and Yan, Qifa and Tang, Xiaohu and Han, Guojun},
	journal={IEEE/ACM Transactions on Networking},
	volume={30},
	number={2},
	pages={717--734},
	year={2021},
	publisher={IEEE}
}

@article{wan2021tradeoff,
	title={On the tradeoff between computation and communication costs for distributed linearly separable computation},
	author={Wan, Kai and Sun, Hua and Ji, Mingyue and Caire, Giuseppe},
	journal={IEEE Transactions on Communications},
	volume={69},
	number={11},
	pages={7390--7405},
	year={2021},
	publisher={IEEE}
}

@article{cheng2025novel,
  title={A Novel Coded Computing Approach for Distributed Multi-Task Learning},
  author={Cheng, Minquan and Wang, Yongkang and Zhang, Lingyu and Wu, Youlong},
  journal={arXiv preprint arXiv:2507.18025},
  year={2025}
}

@article{cadambe2009interference,
  title={Interference alignment and the degrees of freedom of wireless $ X $ networks},
  author={Cadambe, Viveck R and Jafar, Syed A},
  journal={IEEE Transactions on Information Theory},
  volume={55},
  number={9},
  pages={3893--3908},
  year={2009},
  publisher={IEEE}
}

@article{cadambe2008interference,
  title={Interference alignment and degrees of freedom of the $ K $-user interference channel},
  author={Cadambe, Viveck R and Jafar, Syed Ali},
  journal={IEEE transactions on information theory},
  volume={54},
  number={8},
  pages={3425--3441},
  year={2008},
  publisher={IEEE}
}

@article{jafar2008degrees,
  title={Degrees of freedom region of the MIMO $ X $ channel},
  author={Jafar, Syed A and Shamai, Shlomo},
  journal={IEEE Transactions on Information Theory},
  volume={54},
  number={1},
  pages={151--170},
  year={2008},
  publisher={IEEE}
}

@article{khalesi2025tessellated,
  title={Tessellated distributed computing},
  author={Khalesi, Ali and Elia, Petros},
  journal={IEEE Transactions on Information Theory},
  year={2025},
  publisher={IEEE}
}

@article{khalesi2023multi,
  title={Multi-user linearly-separable distributed computing},
  author={Khalesi, Ali and Elia, Petros},
  journal={IEEE Transactions on Information Theory},
  volume={69},
  number={10},
  pages={6314--6339},
  year={2023},
  publisher={IEEE}
}

@incollection{hall1987representatives,
  title={On representatives of subsets},
  author={Hall, Philip},
  booktitle={Classic Papers in Combinatorics},
  pages={58--62},
  year={1987},
  publisher={Springer}
}

@inproceedings{khalesi2024perfect,
  title={Perfect Multi-User Distributed Computing},
  author={Khalesi, Ali and Elia, Petros},
  booktitle={2024 IEEE International Symposium on Information Theory (ISIT)},
  pages={1349--1354},
  year={2024},
  organization={IEEE}
}

@article{khalesi2025lossless,
  title={Lossless Tessellated Distributed Computing},
  author={Khalesi, Ali and Elia, Petros},
  journal={Authorea Preprints},
  year={2025},
  publisher={Authorea}
}

@article{namboodiri2025fundamental,
  title={Fundamental Limits of Distributed Computing for Linearly Separable Functions},
  author={Namboodiri, KK and Peter, Elizabath and Malak, Derya and Elia, Petros},
  journal={arXiv preprint arXiv:2509.23447},
  year={2025}
}

@article{lee2017speeding,
  title={Speeding up distributed machine learning using codes},
  author={Lee, Kangwook and Lam, Maximilian and Pedarsani, Ramtin and Papailiopoulos, Dimitris and Ramchandran, Kannan},
  journal={IEEE Transactions on Information Theory},
  volume={64},
  number={3},
  pages={1514--1529},
  year={2017},
  publisher={IEEE}
}

@inproceedings{ramamoorthy2019universally,
  title={Universally decodable matrices for distributed matrix-vector multiplication},
  author={Ramamoorthy, Aditya and Tang, Li and Vontobel, Pascal O},
  booktitle={2019 IEEE International Symposium on Information Theory (ISIT)},
  pages={1777--1781},
  year={2019},
  organization={IEEE}
}

@inproceedings{das2019distributed,
  title={Distributed matrix-vector multiplication: A convolutional coding approach},
  author={Das, Anindya B and Ramamoorthy, Aditya},
  booktitle={2019 IEEE International Symposium on Information Theory (ISIT)},
  pages={3022--3026},
  year={2019},
  organization={IEEE}
}

@article{qiu2024coded,
  title={Coded Distributed computing for resilient, secure and private matrix-vector multiplication in edge-enabled metaverse},
  author={Qiu, Houming and Zhu, Kun and Luong, Nguyen Cong and Niyato, Dusit and Wu, Qiang},
  journal={IEEE Transactions on Cognitive Communications and Networking},
  volume={10},
  number={5},
  pages={1944--1958},
  year={2024},
  publisher={IEEE}
}

@article{yu2017polynomial,
  title={Polynomial codes: an optimal design for high-dimensional coded matrix multiplication},
  author={Yu, Qian and Maddah-Ali, Mohammad and Avestimehr, Salman},
  journal={Advances in Neural Information Processing Systems},
  volume={30},
  year={2017}
}

@inproceedings{son2023coded,
  title={Coded matrix computation with gradient coding},
  author={Son, Kyungrak and Ramamoorthy, Aditya},
  booktitle={2023 IEEE International Symposium on Information Theory (ISIT)},
  pages={2183--2188},
  year={2023},
  organization={IEEE}
}

@article{zhu2024generalized,
  title={Generalized lagrange coded computing: A flexible computation-communication tradeoff for resilient, secure, and private computation},
  author={Zhu, Jinbao and Tang, Hengxuan and Li, Songze and Chang, Yijia},
  journal={IEEE Transactions on Communications},
  year={2024},
  publisher={IEEE}
}

@inproceedings{ye2018communication,
  title={Communication-computation efficient gradient coding},
  author={Ye, Min and Abbe, Emmanuel},
  booktitle={International Conference on Machine Learning},
  pages={5610--5619},
  year={2018},
  organization={PMLR}
}

@inproceedings{rajbhandari2022deepspeed,
  title={Deepspeed-moe: Advancing mixture-of-experts inference and training to power next-generation ai scale},
  author={Rajbhandari, Samyam and Li, Conglong and Yao, Zhewei and Zhang, Minjia and Aminabadi, Reza Yazdani and Awan, Ammar Ahmad and Rasley, Jeff and He, Yuxiong},
  booktitle={International conference on machine learning},
  pages={18332--18346},
  year={2022},
  organization={PMLR}
}

@article{lin2024moe,
  title={Moe-llava: Mixture of experts for large vision-language models},
  author={Lin, Bin and Tang, Zhenyu and Ye, Yang and Cui, Jiaxi and Zhu, Bin and Jin, Peng and Huang, Jinfa and Zhang, Junwu and Pang, Yatian and Ning, Munan and others},
  journal={arXiv preprint arXiv:2401.15947},
  year={2024}
}
\end{document}